\documentclass[p12]{report}
\usepackage{amsmath}
\def\be{\begin{equation}}
\def\en#1{\label{#1}\end{equation}}

\def\d{\dagger}
\def\bar#1{\overline #1}
\def\vr#1{\mid{#1}\rangle}
\def\vl#1{\langle{#1}\mid}
\def\p{\partial}
\def\l{{\ell}}
\def\vare{\varepsilon}
\def\rd{\mathrm{d}}
\def\br{\mathbf{r}}
\def\bp{\mathbf{p}}
\def\Tr{\mathrm{Tr}}
\def\wtilde#1{\widetilde{#1}}
\def\ba{\mathbf{a}}
\def\bb{\mathbf{b}}

\begin{document}
%%%%%%%%%%%%%%%%%%%%%%%%%%%%%%%%%%%%%%%%%%%%%%%%%%%%%

\pagestyle{empty}

\title{\Huge{The second quantization method for indistinguishable  particles}}

\author{  \Large{Valery  Shchesnovich} \\[12.5cm]
 Lecture Notes for postgraduates\\[0.5cm]   UFABC, 2010}
\date{ }

\maketitle

\begin{quote}
``All I saw hast kept safe in a written record, here thy worth and eminent
endowments come to proof.'' \\[1cm]
\textit{Dante Alighieri, ``Inferno"}
\end{quote}
\newpage

\pagestyle{plain} \setcounter{page}{3} \tableofcontents

\chapter{Creation and annihilation operators for the system of indistinguishable
particles}
\label{chI}

The properties of the permutation  group  are reviewed.  The projectors on the
symmetric and anti-symmetric subspaces of the Hilbert space of the system of
identical particles are considered. The dimension of the physical Hilbert space of
a system of indistinguishable identical particles is calculated. The creation and
annihilation operators are introduced starting from the states they create or
annihilate. The properties of the creation and annihilation operators are derived
from their definition. The basis of the physical Hilbert space is expressed using
the creation operators applied to the vacuum state.  The representation of the
arbitrary vectors in the physical Hilbert space is given in terms of these basis
states. It is shown how to expand the  $s$-particle observables in the operator
basis composed of  the creation and annihilation operators, which corresponds to
the   ``bra-ket'' vector basis of the observables describing distinguishable
particles. The evolution equation for the creation and annihilation operators is
derived. The statistical operators  are also introduced and related to the average
of the observables. Some examples are provided.

\newpage
\section{The permutation  group and the states of a system of indistinguishable  particles}
\label{secI1}

We will say that the particles are identical if they have the same properties. On
the other hand, the particles are considered indistinguishable if they cannot be
labelled with an index.

The Hilbert space of a system consisting of $N$ identical (distinguishable!)
particles is constructed from the tensor product of the states describing each
particle alone, for instance if $N$ particles are found in the states labelled as
$X_\alpha$, $X_\beta$, ..., $X_\gamma$ we have
\be
\vr{X^{(1)}_\alpha,X^{(2)}_\beta,\ldots,X^{(N)}_\gamma} \equiv
\vr{X_\alpha}\vr{X_\beta}\cdot\ldots\cdot\vr{X_\gamma}.
\en{EQ1}
Here and below we will use the upper index for labelling  of the particles, while
in the tensor product (the r.h.s. of Eq. (\ref{EQ1})) the position of the vector
identifies to which particle it corresponds. The arbitrary vector of the Hilbert
space of $N$ identical particles is then given by an arbitrary linear combination
of the vectors defined  in Eq. (\ref{EQ1}).

A permutation  $P$ of $N$ ordered elements $(X_1,X_2,\ldots,X_N)$ is an operation
which changes the order of the elements. A good way to visualize it is to think
that the elements are placed in the string of boxes labelled by the natural numbers
and the operation $P$ swaps the contents of the boxes. For instance, we can write
\be
(X_1,X_2,\ldots,X_N) \xrightarrow{P} (X_{i_1},X_{i_2},\ldots,X_{i_N}),
\en{EQ2}
i.e. implying that $i_k \to  k$ or $P(i_k) = k$. In other words, under the action
of $P$ of Eq. (\ref{EQ2}) the content  from the   $i_k$-th box goes to the box
labelled with the index $k$ (while the content of the $k$-th box goes somewhere
else). The permutation operations form a group, since the composition of two
permutations is also a permutation  and each permutation has the inverse one. The
trivial permutation, which leaves the order unchanged, will be defined by $I$. The
simplest permutation operation is the transposition, i.e. the interchange of just
two elements from the ordered set, for example for $i\ne j$ we define $P_{ij}(i) =
j$ and $P_{ij}(j) = i$ with the rest of the boxes preserving their content under
$P_{ij}$. This permutation  will be denoted below as simply $(i,j)$. Note that it
is inverse to itself $(i,j)(i,j) = I$.  The following useful property
\be
(i,j) = (i,1)(j,2)(1,2)(i,1)(j,2)
\en{EQ3}
can be verified by simple calculation. Moreover, it is easy to see that an
arbitrary permutation  can be represented as composition of the transpositions
(moreover, any permutation  can be written as a composition of the elementary
transpositions of the form $(i,i+1)$). Finally, there are $N!$ of all permutations
of a set of $N$ elements.

The signature of a permutation, denoted as $\vare(P)$, is defined as a mapping of
the permutation  group to the group of two elements $\{-1, 1\}$, where the usual
product is the group operation on the latter set. Using Eq. (\ref{EQ3}) and the
group property one can establish that there are just two possible ways to attribute
a signature   to the permutation: the trivial one, i.e. $P \to 1$ and the one which
attributes  the signature  $-1$ to the permutation   $(1,2)$  and, by Eq.
(\ref{EQ3}),  to any permutation being a transposition, $\vare(P_{ij}) = -1$ for
$i\ne j$ (in this case, the signature is called the parity). To verify that it is
indeed possible to define the parity of $P$ (as preserving the group operation
mapping to the set $\{-1,1\}$) one can produce its explicit expression, given as a
product of the partial signatures, i.e.
\be
\vare(P) \equiv \prod_{i<j}\mathrm{sgn}\left[P(j)-P(i)\right],
\en{EQ4}
where $\mathrm{sgn}$ is the sign function and  the permutation  $P$ acts on the
ordered set $(1,2,\ldots,N)$, e.g.  as in Eq. (\ref{EQ2}).  First, to verify the
group property consider the composition of two permutations $P_2$ and $P_1$, i.e.
the permutation $P_3(k) = P_2(P_1(k))$ or $P_3 = P_2\circ P_1$. We have
\[
\vare(P_2\circ P_1) = \prod_{i<j}\mathrm{sgn}\left[P_2(P_1(j))-P_2(P_1(i))\right]
\]
\[
=\prod_{i<j}\mathrm{sgn}\left[P_1(j)-P_1(i)\right]\prod_{P_1(i)<P_1(j)}\mathrm{sgn}\left[P_2(P_1(j))-P_2(P_1(i))\right]
\]
\[
=\vare(P_1)\vare(P_2).
\]
Second, obviously  $\vare(P_{12}) = -1$. It is clear that the parity $\vare(P) =
(-1)^s$, where $s$ is the number of the transpositions in a representation of $P$
(as a byproduct,  we get that any two such representations must have the same
parity of the number of transpositions).

Let us now consider the permutation  operation  as acting on the Hilbert space of
$N$ identical particles. Since it is a linear space, the permutation  operation $P$
now  becomes a linear operator (for which we will use the same notation $P$).
Namely, we define the action of the operator $P$ corresponding to the permutation
in Eq. (\ref{EQ2})  as follows
\be
P\vr{X^{(1)}_1,X^{(2)}_2,\ldots,X^{(N)}_N} \equiv
\vr{X^{(1)}_{i_1},X^{(2)}_{i_2},\ldots,X^{(N)}_{i_N}},
\en{EQ5}
or in the product form
\[
P\vr{X_1}\vr{X_2}\cdot\ldots\cdot\vr{X_N} =
\vr{X_{i_1}}\vr{X_{i_2}}\cdot\ldots\cdot\vr{X_{i_N}}.
\]
It is clear from this definition that the permutation  operator is unitary, i.e.
$P^\d = P^{-1}$.

If the  particles are considered as indistinguishable, the interchange of any two
must not affect the state of the system (except for a  constant phase). Thus if
$\vr{S}$ is the state of such a system, we must have $P\vr{S} =
e^{i\varphi(P)}\vr{S}$, where the phase depends on the permutation $P$. On the
other hand, for any transposition $P_{ij} = (1,2)$ we must have $P^2_{ij}\vr{S} =
\vr{S}$, since $P_{ij}^2 = I$. Thus $\varphi(P_{ij}) = 0$ or $\pi$ and, due to Eq.
(\ref{EQ3}), it is \textit{the same} phase for any transposition. In other words,
one should use the state which are either symmetric or anti-symmetric in the
interchange of two particles. The particles which are described by the symmetric
states are called \textit{bosons}  and the ones described by the anti-symmetric
states are called \textit{fermions}.  The states which satisfy the property
$P\vr{S} = \vare(P)\vr{S}$, i.e. symmetric and anti-symmetric states, will be
called the physical states of the indistinguishable particles, or simply the
physical states.

To construct a physical state from a product state of $N$ identical particles one
can use the projector operators $S_N$  and $A_N$ defined as follows
\be
S_N \equiv \frac{1}{N!}\sum_P P, \quad A_N \equiv \frac{1}{N!}\sum_P \vare(P)P,
\en{EQ6}
where the summation runs over all transpositions of the set of $N$ elements (note
that in the sum we have operators, while in the index of the summation the
corresponding  transpositions). The projector property is readily verified:
\[
S_N^2 = \left(\frac{1}{N!}\right)^2\sum_P P \sum_{P^\prime} P^\prime =
\left(\frac{1}{N!}\right)^2\sum_P \sum_{P\circ P^\prime}PP^\prime
\]
\[
=\frac{1}{N!}\sum_{P\circ P^\prime}PP^\prime = S_N
\]
and
\[
A_N^2 = \left(\frac{1}{N!}\right)^2\sum_P \vare(P)P \sum_{P^\prime}
\vare(P^\prime)P^\prime = \left(\frac{1}{N!}\right)^2\sum_P \sum_{P\circ
P^\prime}\vare(P)\vare(P^\prime)PP^\prime
\]
\[
=\frac{1}{N!}\sum_{P\circ P^\prime}\vare(PP^\prime)PP^\prime = A_N.
\]
Moreover, the  two projectors in Eq. (\ref{EQ6})  are orthogonal, i.e. the
symmetric and anti-symmetric subspaces are orthogonal (which is a necessary
property if they are to describe two different types of particles, bosons and
fermions). Indeed, by similar calculation one gets
\[
S_N A_N = \left(\frac{1}{N!}\right)^2\sum_P P \sum_{P^\prime}
\vare(P^\prime)P^\prime = \left(\frac{1}{N!}\right)^2\sum_P \vare(P)\sum_{P\circ
P^\prime}\vare(P)\vare(P^\prime)PP^\prime
\]
\[
=\left(\frac{1}{N!}\sum_P\vare(P)\right)A_N = 0,
\]
since one can divide all the  permutation  operators into two classes by
multiplication by $P_{12}$: $\tilde{P} = P_{12} P$, where the operators related by
$P_{12}$ have the signatures of different sign.

The projectors $S_N$ and $A_N$ are also  Hermitian, which is a consequence of the
unitarity of the permutation operators, $P^\d=P^{-1}$  and the property $\vare(P) =
\vare (P^{-1})$.  It is easy to see that only in the case of just two particles the
sum of these projectors is the identity operator, i.e. $S_2+A_2 =I_2$.

To unify the consideration, we will use the same notation $\vare(P)$ for the
signature  in the boson and fermion cases. Thus we will write the ``generalized
symmetrization'' as
\be
S^{\vare}_N  = \frac{1}{N!}\sum_P \vare(P)P,
\en{EQ7}
where $\vare(P) = 1$ for the case of  $S_N$.

The states of $N$ indistinguishable particles   can be given as linear combinations
of the following states
\[
\vr{X_1,X_2,\ldots,X_N} = S^{\vare}_N\{\vr{X^{(1)}_1,X^{(2)}_2,\ldots,X^{(N)}_N}\}
\]
\be
=S^{\vare}_N\{\vr{X_{1}}\vr{X_{2}}\cdot\ldots\cdot\vr{X_{N}}\},
\en{EQ8}
where (and below) we  use the notation $\vr{X_1,X_2,\ldots,X_N}$ for a symmetric or
anti-symmetric state of $N$ particles occupying the single-particle states labelled
by $X_1$, $X_2$, $\ldots$, $X_N$ (not necessarily different). Explicitly, we have
(compare with Eq. (\ref{EQ2}))
\be
\vr{X_1,X_2,\ldots,X_N} =
\frac{1}{N!}\sum_P\vare(P)\vr{X_{P^{-1}(1)}}\vr{X_{P^{-1}(2)}}\cdot\ldots\cdot\vr{X_{P^{-1}(N)}}
\en{EQ9}
Note that this vector is not normalized, in general. If two single-particle states
are identical, $X_i = X_j$ then the corresponding anti-symmetric state is absent
(one obtains zero on the r.h.s. of Eq. (\ref{EQ9})) -- this is nothing but  the
Pauli exclusion principle for  fermions.

%%%%%%%%%%%%%%%%%%%%%%%%%%%%%%%%%%%%%%%%%%%%%%%%%%%%%%%%%%%%%%%%%%%%%%%%%%%%%%%%%%%%%%%%%%%%%%%%%%%%%%%%%%%%%%%%%%%%%%%%%%%%%%

\section{Dimension of the  Hilbert space of a system of indistinguishable  particles}
\label{secI2}

In the previous section we have considered the states of a system of $N$
indistinguishable particles. Consider now the  Hilbert space of such states, the
physical Hilbert space for below. It is a subspace of  the product of  $N$
identical single-particle Hilbert spaces, $H\otimes H\otimes H\otimes \ldots
\otimes H$. We will denote the Hilbert space of $N$ indistinguishable particles as
$\mathcal{H}_N \equiv S^\vare_N\{H\otimes H\otimes H\otimes \ldots \otimes H\}$.
Assuming that  the dimension of $H$ be $d$, i.e. $\mathrm{dim}(H) = d$,  let us
find the dimension of the Hilbert space $\mathcal{H}_N$. The dimension of the
Hilbert space $\mathcal{H}^{(B)}_N$ for bosons, occupying $d$ single-particle
states is the  total amount  of various distributions of the $N$ indistinguishable
particles between the $d$ states, such that there are exactly $m_j$ particles in
the $j$-th state, i.e. (using the usual $\delta$-symbol)
\[
\mathrm{dim}(\mathcal{H}^{(B)}_N) = \sum_{m_1=0}^N\ldots \sum_{m_d=0}^N
\delta_{\sum_{j=1}^dm_j,N}
%\]
%\[
=\sum_{m_1=0}^N\ldots \sum_{m_d=0}^N \int\limits_0^{2\pi}\frac{d\theta}{2\pi}
e^{i\theta\left(\sum_{j=1}^dm_j - N\right)}
\]
\[
=\sum_{m_1=0}^N\ldots \sum_{m_d=0}^N \int\limits_0^{2\pi}\frac{d\theta}{2\pi}
e^{-iN\theta}\prod_{j=1}^de^{im_j\theta} =\int\limits_0^{2\pi}\frac{d\theta}{2\pi}
e^{-iN\theta}\left(\sum_{m=0}^N e^{im\theta}\right)^d
\]
\[
=\int\limits_0^{2\pi}\frac{d\theta}{2\pi}
e^{-iN\theta}\left(\frac{1-e^{i(N+1)\theta}}{1-e^{i\theta}}\right)^d.
\]
Now we change the variable to $z = e^{i\theta}$ and use the theorem of residues  to
calculate the integral (there is a pole at $z=0$, while at $z=1$ the integrand is
regular)
\[
\mathrm{dim}(\mathcal{H}^{(B)}_N) = \frac{1}{2i\pi}\int\limits_{|z|=1} dz\,
z^{-N-1}\left(\frac{1-z^{N+1}}{1-z}\right)^d =
\left.\frac{1}{N!}\frac{d^N}{dz^N}\left(\frac{1-z^{N+1}}{1-z}\right)^d\right|_{z=0}
\]
\[
=\frac{1}{N!}d(d+1)\cdot\ldots\cdot(d+N-1).
\]
Therefore, for bosons we have obtained
\be
\mathrm{dim}(\mathcal{H}^{(B)}_N) =\frac{1}{N!}d(d+1)\cdot\ldots\cdot(d+N-1) =
C_{N+d-1}^N,
\en{EQ10}
where the symbol $C_m^n$ is defined as $C_m^n \equiv \frac{m!}{(m-n)!n!}$.

On the other hand, in the case of fermions by applying the Pauli exclusion
principle (and the indistinguishability of the particles),  we get
\be
\mathrm{dim}(\mathcal{H}^{(F)}_N) = \left\{ \begin{array}{cc}0,& N>d\\
\frac{1}{N!}d(d-1)\cdot\ldots\cdot(d-N+1),& N\le d \end{array} \right. = C_d^N.
\en{EQ11}
%%%%%%%%%%%%%%%%%%%%%%%%%%%%%%%%%%%%%%%%%%%%%%%%%%%%%%%%%%%%%%%%%%%%%%%%%%%%%%%%%%%%%%%%%%%%%%%%%%%

\section{Definition and properties of the creation and annihilation operators}
\label{secI3}

The creation $a^+_\beta$ and annihilation $a^-_\beta$ operators, which
``create''/``annihilate''\footnote{This is exactly what their action means in the
Fock space, see below.} the single-particle state $\beta$, while acting on the
state of $N$ indistinguishable particles, are defined by the following rules
\be
a^-_\beta\vr{\alpha_1,\alpha_2,\ldots,\alpha_N} = \sqrt{N}\langle
\beta^{(1)}\vr{\alpha_1,\alpha_2,\ldots,\alpha_N},
\en{EQ12}
\be
a^+_\beta\vr{\alpha_1,\alpha_2,\ldots,\alpha_N} = \sqrt{N+1}S^\vare_{N+1}\{
\vr{\beta}\vr{\alpha_1,\alpha_2,\ldots,\alpha_N}\},
\en{EQ13}
where the state $\vr{\alpha_1,\alpha_2,\ldots,\alpha_N}$ is the $N$-particle state
defined  as in Eq. (\ref{EQ9}). Thus, in Eq. (\ref{EQ12}) the product is understood
as the scalar product in the  single-particle Hilbert space with the index $i=1$
when the state of the indistinguishable particles is written as a linear
combination of the states of identical particles, Eq. (\ref{EQ9}). Similar, the
action of the projector $S^\vare_{N+1}$ is defined by employing the form of the
state $\vr{\alpha_1,\alpha_2,\ldots,\alpha_N}$ given by Eq. (\ref{EQ9}) (where the
single-particle state $\vr{\beta}$ occupies now the place with index $i=1$, the
state $\vr{\alpha_1}$ the place with index $i=2$ and so on). We will frequently
drop the superscript ``$\vare$'' where it does not lead to a confusion.

Let us first study the action of the annihilation operator. First, we expand the
symmetrization operator $S_N$  by employing   the  permutation, denoted below by
$P_\l$, which brings the $\l$-th element of the ordered set $(1,2,\ldots, N)$, to
the first place while leaving the rest in the same order, i.e. acting as
\be
P_\l(1,2,\ldots, N) = (\l,1,2,\ldots,\l-1,\l+1,\ldots, N).
\en{EQ14}
It has the signature equal to $\xi^{\l-1}$, where $\xi =1$ in the case of bosons
and $\xi=-1$ in the case of fermions. We have the following property
\be
S_N = \frac{1}{N}\sum_{\l=1}^N\xi^{\l-1}S^{[2,\ldots,N]}_{N-1}P_\l,
\en{EQ15}
where the  projector  $S^{[2,\ldots,N]}_{N-1}$ acts on the tensor product of the
$N-1$ single-particle states  with the index  varying from $2$ to $N$.  Thus we
have
\[
a^-_\beta\vr{\alpha_1,\alpha_2,\ldots,\alpha_N} =
\frac{1}{\sqrt{N}}\sum_{\l=1}^N\xi^{\l-1}\langle\beta\vr{\alpha_\l}
\underbrace{\vr{\alpha_1,\alpha_2,\ldots,\alpha_N}}_{\vee\alpha_\l}
\]
\be
\equiv\frac{1}{\sqrt{N}}\sum_{\l=1}^N\xi^{\l-1}\langle\beta\vr{\alpha_\l}S_{N-1}\{
\vr{\alpha_1}\cdot\ldots\cdot\vr{\alpha_{\l-1}}\vr{\alpha_{\l+1}}\cdot\ldots\cdot\vr{\alpha_N}\}.
\en{EQ16}
Here and below the  underbrace means that the state labelled by $\alpha_\l$ is
omitted from the expression for the $N$-particle state
\mbox{$\vr{\alpha_1,\alpha_2,\ldots,\alpha_N}$} making it a $(N-1)$-particle state
(also of indistinguishable particles in this case).

On the other hand, using the obvious property $S_{N+1}S^{[2,\ldots,N+1]}_{N} =
S_{N+1}$ (since $S^{[2,\ldots,N+1]}_{N}$ projects on the subspace of the projector
$S_{N+1}$) we obtain from Eq. (\ref{EQ13}) the action of the creation operator as
follows
\be
a^+_\beta\vr{\alpha_1,\alpha_2,\ldots,\alpha_N} = \sqrt{N+1}S^\vare_{N+1}\{
\vr{\beta}\vr{\alpha_1}\vr{\alpha_2}\cdot\ldots\cdot\vr{\alpha_N}\}.
\en{EQ17}

Note that expressions (\ref{EQ16}) and (\ref{EQ17}) make it evident that the
resulting  states belong to the respective physical Hilbert spaces, i.e. to
$\mathcal{H}_{N-1}$ and $\mathcal{H}_{N+1}$, respectively.

Another useful representation of the action of the creation operator, which is used
below, is given by expanding the symmetrization operator  similar to Eq.
(\ref{EQ15}),  but using now  the inverse operator $P^{-1}_\l$ instead, i.e. which
acts as
\[
P^{-1}_\l(1,2,\ldots, N) = (2,\ldots,\l-1,\l,1,\l+1,\ldots, N),
\]
where $P^{-1}_1 = I$ since $P_1 = I$.  With the  use of $P_\l^{-1}$ the
$(N+1)$-particle symmetrization operator is cast as
\[
S_{N+1} = \frac{1}{N+1}\sum_{\l=1}^{N+1}\xi^{\l-1}S^{[2,\ldots,N+1]}_{N}P^{-1}_\l.
\]
Using this property, we get
\[
a^+_\beta\vr{\alpha_1,\alpha_2,\ldots,\alpha_N} =
\frac{1}{\sqrt{N+1}}\biggl[\vr{\beta^{(1)}}\vr{\alpha^{(2)}_1,\ldots,\alpha^{(N+1)}_N}
\]
\be
+\sum_{\l=1}^{N}\xi^{\l}\vr{\alpha^{(1)}_\l}S_N
\{\underbrace{\vr{\beta^{(2)},\alpha^{(3)}_1,\ldots,\alpha^{(N+1)}_N}}_{\vee\alpha_\l}\}\biggr].
\en{EQ18}
Note that this expression resembles some features of that for the annihilation
operator, Eq. (\ref{EQ16}).

Now, using the definitions of the creation and annihilation operators one can
easily prove the following properties:
\begin{eqnarray}
&&(a^{\pm}_\beta)^\d = a^{\mp}_\beta,\label{EQ19}\\
&& a^{\pm}_\gamma a^{\pm}_\beta - \xi  a^{\pm}_\beta a^{\pm}_\gamma = 0,\label{EQ20}\\
&& a^{-}_\gamma a^{+}_\beta - \xi  a^{+}_\beta a^{-}_\gamma =
\langle\gamma\vr{\beta},\label{EQ21}
\end{eqnarray}
where $\xi =1$ for bosons and $\xi =-1$ for fermions.

\textit{Proof of the properties} (\ref{EQ19})-(\ref{EQ21}). I. Consider property
(\ref{EQ19}). We have to prove that for any two vectors
\[
\left(\vl{\beta_1,\ldots,\beta_{N+1}}a^+_\gamma\vr{\alpha_1,\ldots,\alpha_{N}}\right)^\d
= \vl{\alpha_1,\ldots,\alpha_{N}}a^-_\gamma\vr{\beta_1,\ldots,\beta_{N+1}}.
\]
Indeed, by the definition of the creation operator, Eq. (\ref{EQ13}), we have
\[
\left(\vl{\beta_1,\ldots,\beta_{N+1}}a^+_\gamma\vr{\alpha_1,\ldots,\alpha_{N}}\right)^\d
\]
\[
=\sqrt{N+1}\left(\vr{\beta_1,\ldots,\beta_{N+1}}S_{N+1}\{\vr{\gamma}\vr{\alpha_1,\ldots,\alpha_{N}}\}\right)^\d
\]
\[
=\sqrt{N+1}\left(\vl{\beta_1,\ldots,\beta_{N+1}}\gamma\rangle\vr{\alpha_1,\ldots,\alpha_{N}}\right)^\d
\]
\[
=\vl{\alpha_1,\ldots,\alpha_{N}}\sqrt{N+1}\langle\gamma\vr{\beta_1,\ldots,\beta_{N+1}}
\]
\[
=\vl{\alpha_1,\ldots,\alpha_{N}}a^-_\gamma\vr{\beta_1,\ldots,\beta_{N+1}},
\]
where we have  used $S^\d_{N+1} = S_{N+1}$, $S_{N+1}^2 = S_{N+1}$  and the
definition of the annihilation operator, Eq. (\ref{EQ12}).

II. Consider now the property (\ref{EQ20}). Due to Eq. (\ref{EQ16}) we have
\[
a^-_\gamma a^-_\beta \vr{\alpha_1,\ldots,\alpha_N} =
a^-_\gamma\left(\frac{1}{\sqrt{N}}\sum_{\l=1}^N\xi^{\l-1}\langle\beta\vr{\alpha_\l}
\underbrace{\vr{\alpha_1,\ldots,\alpha_N}}_{\vee \alpha_\l}\right)
\]
\[
=\frac{1}{\sqrt{N-1}}\frac{1}{\sqrt{N}}
\left(\sum_{\l=1}^N\sum_{m=1}^{\l-1}\xi^{m-1}\xi^{\l-1}\langle\gamma\vr{\alpha_m}\langle\beta\vr{\alpha_\l}
\underbrace{\vr{\alpha_1,\ldots,\alpha_N}}_{\vee\alpha_\l,\vee\alpha_m}\right.
\]
\[
+\left.\sum_{\l=1}^N\sum_{m=\l+1}^{N}\xi^{m-2}\xi^{\l-1}\langle\gamma\vr{\alpha_m}\langle\beta\vr{\alpha_\l}
\underbrace{\vr{\alpha_1,\ldots,\alpha_N}}_{\vee\alpha_\l,\vee\alpha_m} \right),
\]
where in the second and third lines the states labelled $\alpha_\l$ and $\alpha_m$
are omitted from the state vector. By interchanging the indices  $\l$ and $m$ in
the third line we obtain
\[
a^-_\gamma a^-_\beta \vr{\alpha_1,\ldots,\alpha_N} =
\frac{1}{\sqrt{N-1}}\frac{1}{\sqrt{N}}\sum_{\l=1}^N\sum_{m=1}^{\l-1}\xi^{\l+m-2}\biggl(
\langle\gamma\vr{\alpha_m}\langle\beta\vr{\alpha_\l}
\]
\be
+\xi\langle\gamma\vr{\alpha_\l}\langle\beta\vr{\alpha_m}\biggr)
\underbrace{\vr{\alpha_1,\ldots,\alpha_N}}_{\vee\alpha_\l,\vee\alpha_m}.
\en{EQ22}
Property (\ref{EQ20})  for the annihilation operators evidently follows from  Eq.
(\ref{EQ22}). For the creation operators this property follows by application of
property (\ref{EQ19}).

III. Consider the two sides of Eq. (\ref{EQ21}) applied to a state vector. We have
by the definitions, Eqs. (\ref{EQ12}) and (\ref{EQ13}),
\[
a^{-}_\gamma
a^{+}_\beta\vr{\alpha_1,\ldots,\alpha_N}=(N+1)\langle\gamma|S_{N+1}\{\vr{\beta}\vr{\alpha_1,\ldots,\alpha_N}\},
\]
\be
=\langle\gamma\vr{\beta}\vr{\alpha_1,\ldots,\alpha_N}+
\sum_{\l=1}^N\xi\langle\gamma\vr{\alpha_\l}\underbrace{\vr{\alpha_1,\ldots,\alpha_N}}_{\alpha_\l\to\beta},
\en{EQ23}
where we have used the property
\be
S_{N+1} = \frac{1}{N+1}S^{[2,\ldots,N+1]}_{N}\left(I_{N+1} + \sum_{\l=2}^{N+1}\xi
P_{1\l}\right),
\en{EQ24}
where $P_{1\l}$ is the operator corresponding to the permutation  $(1,\l)$. On the
other hand we have
\[ a^{+}_\beta
a^{-}_\gamma\vr{\alpha_1,\ldots,\alpha_N}
=\sum_{\l=1}^N\xi^{\l-1}\langle\gamma\vr{\alpha_\l}S_N
\{\vr{\beta}\underbrace{\vr{\alpha_1}\cdot\ldots\cdot\vr{\alpha_N}}_{\vee\alpha_\l}\}
\]
\be
=\sum_{\l=1}^N\langle\gamma\vr{\alpha_\l}\underbrace{\vr{\alpha_1,\ldots,\alpha_N}}_{\alpha_\l\to\beta}.
\en{EQ25}
where we have used Eqs. (\ref{EQ13}) and (\ref{EQ16}) (in the last line the
underbrace means the the state labelled by $\alpha_\l$ is replaced by the one
labelled by $\beta$). Comparing expressions (\ref{EQ23}) and (\ref{EQ25}) one
arrives at property (\ref{EQ21}).

Since the creation and annihilation operators are, by Eq. (\ref{EQ19}),  Hermitian
conjugated to each other, we will adopt the notation $a_{\alpha}$  for the
annihilation operator, while the creation operator will be denoted by
$a^\d_{\alpha}$.

One  easily verify the following properties (valid for both types of particles,
bosons and fermions):
\be
[a^\d_\alpha a_\alpha,a^\d_\beta] = \langle\alpha\vr{\beta}a^\d_\alpha,\quad
[a^\d_\alpha a_\alpha,a_\beta] = -\langle\beta\vr{\alpha}a_\alpha.
\en{EQ27}

An important property of the creation/annihilation operators, which is a direct
consequence of their definitions, Eqs. (\ref{EQ12}) and (\ref{EQ13}), is that they
depend  linearly on the state which they ``create''/``annihilate'' (more precisely,
the creation operator is linear in the state and the annihilation operator is
anti-linear in the state, the latter is linear in the Hermitian conjugated state).
Writing explicitly the dependence of the creation/annihilation operator on the
state, i.e.
\be
a^\d_\alpha \equiv A^+(\vr{\alpha}),\quad a_\alpha \equiv A^-(\vl{\alpha}),
\en{EQ28}
we have the following linearity property
\be
A^+\biggl(\sum_jC_j\vr{X_j}\biggr) = \sum_jC_jA^+\bigl(\vr{X_j}\bigr), \;
A^-\biggl(\sum_jD_j\vl{X_j}\biggr) = \sum_jD_jA^-\bigl(\vl{X_j}\bigr).
\en{EQ29}
This property allows the change of basis for the creation and annihilations
operators, quite similarly as the change of basis is performed for the quantum
states themselves. Indeed, expanding the state $\vr{\alpha}$ in a basis
$\vr{\alpha} = \sum_j\vr{j}\langle j\vr{\alpha}$ we get
\be
a^\d_\alpha = \sum_j\langle j\vr{\alpha}a^\d_j,\quad a_\alpha = \sum_j\langle
\alpha\vr{j}a_j.
\en{EQ30}

%%%%%%%%%%%%%%%%%%%%%%%%%%%%%%%%%%%%%%%%%%%%%%%%%%%%%%%%%%%%%%%%%%%%%%%%%%%%%%%%%%%%%%%%%%%%%%%%%%%%%%%%%%%%%%%%%%%%%%%%%%%%%%%

\section{The Fock space }
\label{secI4}

The linearity of the creation and annihilation operators on the ``created'' or
``annihilated'' state implies that it is sufficient to consider them for the
orthogonal basis states. More importantly, their transformations between the basis
states follow from those of the basis states themselves, as in  Eq. (\ref{EQ30}).
Only the number of particles in each single-particle state, comprising the
$N$-particle state, is the most  important parameter.  Thus we have naturally come
to the concept of the occupation number representation of the physical Hilbert
space, i.e. to the concept of the Fock space. Suppose that we are given a basis in
the Hilbert space $H$ of the single-particle states
\be
\vr{1}, \vr{2}, \vr{3}, \ldots, \vr{k},\ldots; \quad \langle k\vr{j} = \delta_{kj}.
\en{EQ31}
It is convenient to use the ``variable index notation'' for the basis states, which
does not specify which particular $N$ states of the   basis   are selected, i.e.
for $N$ particles we will write the basis state as $\vr{k_j}$, with
$j=1,2,\ldots,N$ (here $k_j$ is a natural number and some $k_j$ with different $j$
may coincide). The $N$-particle states for indistinguishable particles are given
now as
\be
\vr{k_1,k_2,\ldots,k_N}
=\frac{1}{N!}\sum_P\vare(P)\vr{k_{P^{-1}(1)}}\cdot\ldots\cdot\vr{k_{P^{-1}(N)}}.
\en{EQ32}
The state defined by Eq. (\ref{EQ32}) is, in general,  not normalized.  One can
define the  normalized $N$-particle states for the indistinguishable particles,
which we will call the Fock states. For bosons  the normalized states are given as
\be
\vr{n_{k_{j_1}},n_{k_{j_2}},\ldots,n_{k_{j_s}}} \equiv \left(
\frac{N!}{n_{k_{j_1}}!n_{k_{j_2}}!\cdot\ldots\cdot
n_{k_{j_s}}!}\right)^{\frac{1}{2}}\vr{k_1,k_2,\ldots,k_N},
\en{EQ33}
where we assume that only the states $k_j$, $j=1,\ldots, s$ are occupied,  while
for fermions we can also write\footnote{Due to the asymmetry with respect to the
labels $k_j$ the fermion $N$-particle state depends on the order of the
single-particle states in it, a swap of any two such states leads to a
$\pi$-phase.}
\be
\vr{n_{1},n_{2},\ldots,n_{N}}\equiv \sqrt{N!}\vr{k_1,k_2,\ldots,k_N},
\en{EQ34}
where due to the Pauli exclusion principle all $n_{j}=1$.  The numbers $n_j$ are
called the occupation numbers (for fermions the occupation numbers $n_j$ take just
two values $0$ and $1$). The normalizing factors in Eqs. (\ref{EQ33}) and
(\ref{EQ34}) follow from the definition of the state $\vr{k_1,k_2,\ldots,k_N}$, Eq.
(\ref{EQ32}), and simple calculation.

Let us consider the action of the creation and annihilation operators on the Fock
states (\ref{EQ33}) and (\ref{EQ34}).

\textit{I. Boson case}. We have (for simplicity, we rename $n_{k_j}$ into $n_k$)
\[
a_{k_j}\vr{n_{k_1},\ldots,n_{k_s}} =\left( \frac{N!}{n_{k_1}!\cdot\ldots\cdot
n_{k_s}!}\right)^{\frac{1}{2}}a_{k_j}\vr{k_1,k_2,\ldots,k_N}
\]
\[
=\left( \frac{N!}{n_{k_1}!\cdot\ldots\cdot
n_{k_s}!}\right)^{\frac{1}{2}}\frac{1}{\sqrt{N}}\sum_{l=1}^N\langle
k_j\vr{k_l}\underbrace{\vr{k_1,\ldots,k_N}}_{\vee k_l}
\]
\[
=\left( \frac{(N-1)!}{n_{k_1}!\cdot\ldots\cdot
n_{k_s}!}\right)^{\frac{1}{2}}\sum_{k_l=k_j}\underbrace{\vr{k_1,\ldots,k_N}}_{\vee
k_l}
\]
\be
=\sqrt{n_{k_j}}\vr{n_{k_1},\ldots,n_{k_j}-1,\ldots,n_{k_s}},
\en{EQ35}
where we have used Eq. (\ref{EQ16}) and the orthogonality of the single-particle
basis vectors (the last summation is  over all indices $k_l$ coinciding with
$k_j$). Similarly
\[
a^\d_{k_j}\vr{n_{k_1},\ldots,n_{k_s}} =\left( \frac{N!}{n_{k_1}!\cdot\ldots\cdot
n_{k_s}!}\right)^{\frac{1}{2}}\sqrt{N+1}S_{N+1}\{\vr{k_j}\vr{k_1,k_2,\ldots,k_N}\}
\]
\[
=\left( \frac{(N+1)!}{n_{k_1}!\cdot\ldots\cdot (n_{k_j}+1)!\cdot\ldots\cdot
n_{k_s}!}\right)^{\frac{1}{2}}\sqrt{n_{k_j}+1}\vr{k_j,k_1,k_2,\ldots,k_N}
\]
\be
=\sqrt{n_{k_j}+1}\vr{n_{k_1},\ldots,n_{k_j}+1,\ldots,n_{k_s}},
\en{EQ36}
where we have used the definition, Eq. (\ref{EQ13}), and the  symmetry property of
the state vector for bosons.

Let us also introduce the vacuum state, $\vr{Vac}$, i.e. the state with no
particles ($N=0$). It has the following property
\be
a_{k_j}\vr{Vac} = 0,
\en{EQ37}
for any annihilation operator $a_{k_j}$ (and, due to the linearity property of the
operator in the state it annihilates, Eq. (\ref{EQ29}),  for arbitrary boson
annihilation operator). Then, one can use Eq. (\ref{EQ35}) to represent the Fock
basis state in the following form
\be
\vr{n_{k_1},n_{k_2},\ldots,n_{k_s}}  = \frac{(a^\d_{k_1})^{n_{k_1}}\cdot\ldots\cdot
(a^\d_{k_s})^{n_{k_s}}}{\sqrt{n_{k_1}!\cdot\ldots\cdot n_{k_s}!}}\vr{Vac}.
\en{EQ38}

\textit{II. Fermion case}. In the fermion case, as different from the boson case,
the order of the single-particle states in the $N$-particle state is important.
Thus, the formulae derived above for bosons have their analogs for fermions with
the additional phases due to transpositions used to add or remove the
single-particle states to and from the $N$-particle one. Using Eq. (\ref{EQ16}) we
have, for instance,
\[
a_{k_j}\vr{n_{k_1},n_{k_2},\ldots,n_{k_N}} =
\frac{1}{\sqrt{N}}\sum_{\l=1}^N(-1)^{\sum_{m<j} n_{k_m}}\langle
k_j\vr{k_\l}\underbrace{\vr{k_1,\ldots,k_N}}_{\vee k_\l}
\]
\[
=\delta_{n_{k_j},1}(-1)^{\sum_{l<j} n_{k_l}}\underbrace{\vr{k_1,\ldots,k_N}}_{\vee
k_j}
\]
\be
=(-1)^{\sum_{l<j}
n_{k_l}}\sqrt{n_{k_j}}\vr{n_{k_1},\ldots,n_{k_j}-1,\ldots,n_{k_N}},
\en{EQ39}
where $n_{k_m}$ is the occupation number of the $k_m$ state. Similarly,
\[
a^\d_{k_j}\vr{n_{k_1},n_{k_2},\ldots,n_{k_N}}  =
\sqrt{N+1}A_{N+1}\{\vr{k_j}\vr{n_{k_1},n_{k_2},\ldots,n_{k_N}}
\]
\[
=\sqrt{(N+1)!}A_{N+1}\{\vr{k_j}\vr{k_1}\cdot\ldots\cdot\vr{k_N}\}
\]
\[
=\delta_{n_{k_j},0}(-1)^{\sum_{l<j}
n_{k_l}}\sqrt{(N+1)!}A_{N+1}\{\vr{k_1}\cdot\ldots\cdot\vr{k_j}\cdot\ldots\cdot\vr{k_N}\}
\]
\be
=(-1)^{\sum_{l<j}
n_{k_l}}\sqrt{n_{k_j}+1}\vr{n_{k_1},\ldots,n_{k_j}+1,\ldots,n_{k_N}},
\en{EQ40}
where we have used that $A_{N+1}A_{N}^{[2\ldots N+1]} = A_{N+1}$ and the fact that
for $n_{k_j}=1$ the state $\vr{n_{k_1},\ldots,n_{k_j}+1,\ldots,n_{k_N}}$ becomes
zero.

In the fermion case the Fock state (with all $n_{k_j}=1$) is given as
\be
\vr{n_{k_1},n_{k_2},\ldots,n_{k_N}}  = \sqrt{N!}\vr{{k_1},{k_2},\ldots,{k_N}}  =
a^\d_{k_1}\cdot\ldots\cdot a^\d_{k_N}\vr{Vac},
\en{EQ41}
where $\vr{Vac}$ is the fermion vacuum state.

For the boson and fermion cases one can define the occupation-number operator as
\be
n_{k} = a^\d_k a_k,
\en{EQ42}
which gives the occupation number of the corresponding single-particle state (this
operator is a scalar in the Fock space, thus we use the same notation).

Note that  one can write down the \textit{unnormalized} $N$-particle basis state of
the indistinguishable particles for both, boson and fermion cases, as follows
\be
\vr{{k_1},{k_2},\ldots,{k_N}} = \frac{1}{\sqrt{N!}} a^\d_{k_1}\cdot\ldots\cdot
a^\d_{k_N}\vr{Vac},
\en{EQ43}
where in the boson case, some $k_j$ with different index $j$ may coincide. The
importance of this representation will become clear below. The expansion of unit
operator, i.e. the projector  $S_N$  on the physical Hilbert space of $N$
indistinguishable particles now can be written as follows (for bosons and fermions)
\be
\sum_{\sum_j n_j = N} \vr{n_{1},n_{2},\ldots,n_{s},\ldots}
\vl{n_{1},n_{2},\ldots,n_{s},\ldots} = S_N,
\en{EQ44}
where the summation is over all sets of occupation numbers, such that their sum is
$N$.  Moreover, due to the summation identity
\be
\sum_{\sum_j n_j = N}\frac{N!}{n_{k_1}!n_{k_2}!\cdot\ldots\cdot n_{k_s}!}\{\ldots\}
=\sum_{k_1} \ldots \sum_{k_N}\{\ldots\},
\en{SUMID}
the same projector  operator can be expressed using the unnormalized states as
\be
\sum_{k_1} \ldots \sum_{k_N}\vr{k_{1},k_{2},\ldots,k_{N}}
\vl{k_{1},k_{2},\ldots,k_{N}} = S_N,
\en{EQ45}
where  the last formula has an important advantage for below, since  the summations
over the state indices $k_j$  vary \textit{independently} of each other.
%%%%%%%%%%%%%%%%%%%%%%%%%%%%%%%%%%%%%%%%%%%%%%%%%%%%%%%%%%%%%%%%%%%%%%%%%%%%%%%%%%%%%%%%%%%%%%%%%

\section{The representations of state vectors and  operators}
\label{secI5}

\subsection{$N$-particle wave-functions}
We did not specify the basis for construction of the Fock space and used only the
orthogonality of the single-particle basis states. In this respect, the linearity
of the creation and annihilation operators in the state they create or annihilate
allows to pass from one Fock space to the other. For instance, assume that we pass
from the basis of infinite but countable number states to the co-ordinate
representation as follows (here the index $k$ may include also the spin variable)
\be
\vr{k} = \int \rd x\varphi_k(x)\vr{x},
\en{EQ46}
where $\varphi_k(x)\equiv \langle x\vr{k}$ is the wave-function corresponding to
the basis state $\vr{k}$ and $\vr{x}$ is the basis state in the co-ordinate
representation. Denoting the creation and annihilation operators of the co-ordinate
basis state $\vr{x}$ as $\psi^\d(x)$ and $\psi(x)$ by Eq. (\ref{EQ29}) we have
\be
a^\d_k = \int \rd x\varphi_k(x)\psi^\d(x),  \quad a_k = \int \rd
x\varphi^*_k(x)\psi(x).
\en{47}
The inverse transformation also follows from Eq. (\ref{EQ29}):
\be
\psi(x) = \sum_k \varphi_k(x) a_k,\quad  \psi^\d(x) = \sum_k \varphi^*_k(x) a^\d_k.
\en{EQ48}

In the same way as for the one-particle case, one can define the wave-function for
the case of $N$ indistinguishable particles. Indeed, one just need to project the
Fock state (a multi-particle analog of the basis state $\vr{k}$) onto the expansion
(\ref{EQ45}) of the unit operator in the physical space, i.e. the projector $S_N$.
Using the basis states from the co-ordinate representation  (with the summation in
Eq. (\ref{EQ45}) substituted by the integration over all coordinates) we define
\be
\varphi(x_1,\ldots,x_N) \equiv \langle x_1, \dots, x_N\vr{n_{k_{j_1}},\ldots,
n_{k_{j_s}}},
\en{EQ49}
so that now
\[
\vr{n_{k_{j_1}},\ldots, n_{k_{j_s}}} =\int \rd x_1\ldots \int \rd x_N
\varphi(x_1,\ldots,x_N)\vr{x_1,\ldots,x_N}
\]
\be
=\frac{1}{\sqrt{N!}}\int \rd x_1\ldots \int \rd x_N \varphi(x_1,\ldots,x_N)
\psi^\d(x_1)\cdot\ldots\cdot\psi^\d(x_N)\vr{Vac},
\en{EQ50}
where we have used Eq. (\ref{EQ43}) now for the co-ordinate basis states of $N$
indistinguishable particles. Thus one may even use the notation $\vr{\varphi_N}$
for this Fock state. The scalar product between two $N$-particle wave-functions is
given by the usual integral formula, again due to the property (\ref{EQ45}), i.e.
\be
\langle \phi_N\vr{\varphi_N} = \int \rd x_1\ldots \int \rd x_N
\phi^*(x_1,\ldots,x_N)\varphi(x_1,\ldots,x_N),
\en{EQ51}
where the co-ordinates under the integrals vary independently. Hence, we have
arrived at  the ``usual'' $N$-particle wave-function representation for the state
of $N$ indistinguishable particles, which also can be obtained as appropriate
symmetrization of the single-particle wave-functions. Indeed we have from the
definitions (\ref{EQ33}) and (\ref{EQ49})
\[
\varphi(x_1,\ldots,x_N) = \langle x_1, \dots,
x_N|\left(\frac{N!}{n_{k_{j_1}}!n_{k_{j_2}}!\cdot\ldots\cdot
n_{k_{j_s}}!}\right)^{\frac{1}{2}}S_N\{\vr{k_{1}}\cdot\ldots\cdot\vr{k_{N}}\}
\]
\[
=\left(\frac{N!}{n_{k_{j_1}}!n_{k_{j_2}}!\cdot\ldots\cdot
n_{k_{j_s}}!}\right)^{\frac{1}{2}}S_N\{ \langle x_1\vr{k_{1}} \cdot \ldots
\cdot\langle x_N\vr{k_N}\}
\]
\be
=\left(\frac{N!}{n_{k_{j_1}}!n_{k_{j_2}}!\cdot\ldots\cdot
n_{k_{j_s}}!}\right)^{\frac{1}{2}}S_N\{\varphi_{k_1}(x_1)\cdot\ldots\cdot\varphi_{k_N}(x_N)\},
\en{EQ52}
where we have used the projector property $S_N^2 = S_N$ allowing to remove the
symmetrization from the basis vector $\vl{x_1,\ldots,x_N}$ before evaluation  of
the scalar product (\textit{note the normalizing factor in this definition}).

\subsection{The second-quantization  representation  of operators}
\label{subI5A}

The observables of the system of indistinguishable particles, the physical
operators for below,  are  just  the symmetric subset of all operators acting in
the Hilbert space of identical (i.e. distinguishable) particles. Thus the physical
operators can be expanded over the basis of the Hilbert space of identical
particles. However, we will be interested in the representation of these in the
physical subspace only, i.e. how they act  over the space of indistinguishable
particles. We will show that their action can be expressed as action of the
products of the creation and annihilation operators.

\noindent\textit{The one-particle operators.}
Let us start with the one-particle operators (which act on the $N$-particle Hilbert
space and  should not be confused with single-particle ones). We have a general
structure of such an operator
\be
A_1 = A^{(1)} + A^{(2)}+ \ldots + A^{(N)},
\en{EQ53}
where $A^{(j)}$ is the single-particle operator acting on the Hilbert space of the
the $j$th particle. A one-particle operator (\ref{EQ53}) is symmetric as it should
be for an observable describing a system of indistinguishable particles.

If we use two basis vectors of the single-particle Hilbert space, such that the
matrix element of a single-particle operator is $\langle \alpha
|A^{(1)}|\beta\rangle$, the representation of a one-particle operator in the
Hilbert space of identical particles  has the form
\be
A_1 =   \sum_\alpha\sum_\beta\langle \alpha
|A^{(1)}|\beta\rangle\sum_{j=1}^N\vr{\alpha^{(j)}}\vl{\beta^{(j)}},
\en{EQ54}
where the lust summation is understood as a direct sum of the operators acting in
the single-particle Hilbert spaces indexed  from 1 to $N$. The aim is find how a
one-particle operator acts on the physical subspace.   By Eq. (\ref{EQ54}) this
amounts to finding out how the  the last sum acts in the physical subspace. The
latter is given by the following identity
\be
\sum_j\vr{\alpha^{(j)}}\vl{\beta^{(j)}} = a_\alpha^\d a_\beta,
\en{EQ55}
\textit{valid in the physical space}, i.e. the Hilbert space of indistinguishable
particles. Indeed, let us prove it by analyzing how it acts on a $N$-particle state
(of the indistinguishable particles) and comparing to the product of the operators
on the r.h.s.. We have obviously
\[
\sum_{j=1}^N\vr{\alpha^{(j)}}\vl{\beta^{(j)}}k_1,\ldots,k_N\rangle =
\]
\[
=\frac{1}{N}\sum_{j=1}^N\sum_{\l =1}^N\xi^{\l-j}\langle\beta|k_\l\rangle
S^{[1\ldots j-1,j+1\ldots
N]}_{N-1}\{\underbrace{\vr{k_1^{(1)}}\cdot\ldots\cdot\vr{\alpha^{(j)}}\cdot
\ldots\cdot\vr{k_N^{(N)}}}_{\vee k_\l}\}
\]
\[
=\sum_{\l=1}^N\langle\beta|k_\l\rangle\underbrace{\vr{k_1,\ldots,k_N}}_{k_\l\to\alpha},
\]
here we have used the factorization of the symmetrization operator $S_N$ similar to
Eq. (\ref{EQ15}) where instead the $k_\l$th vector is brought not to the first but
to the $j$th place. Comparing with Eq. (\ref{EQ25}) we get the result (\ref{EQ55}).
Note that  the r.h.s. of Eq. (\ref{EQ55}) does not depend on the number of
particles, thus leading to a representation for the many-particle operator $A_1$ in
Eq. (\ref{EQ54}) which is $N$-independent. We have in particular
\be
A_1 =   \sum_\alpha\sum_\beta\langle \alpha |A^{(1)}|\beta\rangle a_\alpha^\d
a_\beta.
\en{EQ56}
Eq. (\ref{EQ55}) can also be shown using the following properties
\[
\xi^{j-1}\sqrt{N}\langle\beta^{(j)}\vr{k_1,\ldots,k_N} =
a_\beta\vr{k_1,\ldots,k_N},
\]
\be
\frac{1}{\sqrt{N}}\sum_{j=1}^N\xi^{j-1}\vr{\alpha^{(j)}}  \vr{k_1,\ldots,k_N} =
a_\alpha^\d  \vr{k_1,\ldots,k_N},
\en{EQ57}
where the first property follows from Eq. (\ref{EQ16}) noticing that
$\langle\beta^{(j)}\vr{k_1,\ldots,k_N} =
\xi^{j-1}\langle\beta^{(1)}\vr{k_1,\ldots,k_N}$, while in the second it is
understood that new vector occupies the $j$th place in the product of the
single-particle vectors in the state \mbox{$\vr{k_1,\ldots,k_N}$} and  we have used
Eq. (\ref{EQ17}) and a factorization of the symmetrization operator similar to that
of Eq. (\ref{EQ15}).
%%%%%%%%%%%%%%%%%%%%%%%%%%%%%%%%%%%%%%%%%%%%%%%%%%%%%%%%%%%%%%%%%%%%%%%%%%%%%%%%%%

\noindent\textit{The many-particle operators.}
The generalization to the $s$-particle operators, acting in the Hilbert space of
indistinguishable particles is obvious. We have
\be
A_s = \sum_{j_1<j_2<\ldots <j_s}A^{(j_1,\ldots,j_s)} =
\frac{1}{s!}\sum_P\sum_{j_1<\ldots<j_s}A^{(P(j_1),\ldots,P(j_s))},
\en{EQ58}
where to satisfy the property of being an observable describing the system of
indistinguishable particles, the operator $A^{(j_1,\ldots,j_s)}$ must be symmetric
with respect to permutations of  the indices $j_1, \ldots, j_s$. Indeed, by
definition of the physical observable, for any permutation $P$ of the state of
identical particles we must have
\[
\vl{f_1^{(1)},\ldots,f_N^{(N)}}A_s\vr{k_1^{(1)},\ldots,k_N^{(N)}} =
\vl{f_1^{(1)},\ldots,f_N^{(N)}}P^\d A_sP \vr{k_1^{(1)},\ldots,k_N^{(N)}},
\]
i.e. $P^\d A_sP = A_s$ and since $P^\d A^{(j_1,\ldots,j_s)}P =
A^{(P^{-1}(j_1),\ldots,P^{-1}(j_s))}$ (compare with Eq. (\ref{EQ9})).

Let us now generalize Eq. (\ref{EQ55}) to the $s$-particle representation. We have
\be
\sum_P\sum_{j_1<\ldots<j_s}\vr{\alpha^{(j_1)}_{P(1)}}\cdot\ldots\cdot\vr{\alpha^{(j_s)}_{P(s)}}
\vl{\beta^{(j_s)}_{P(s)}}\cdot\ldots\cdot\vl{\beta^{(j_1)}_{P(1)}} =
a^\d_{\alpha_1}\cdot\ldots\cdot a^\d_{\alpha_s} a_{\beta_s}\cdot\ldots\cdot
a_{\beta_1},
\en{EQ59}
again, \textit{valid in the physical space}, i.e. the Hilbert space of
indistinguishable particles. To see why  a permutation $P$ must   appear, note that
the summation over the ordered indices $j_1< \ldots<j_s$ is lifted to the summation
over the unequal indices $j_1,\ldots, j_s$ by this permutation, while by Eq.
(\ref{EQ55}) it is almost evident that the indices in the summation can  indeed
appear in an arbitrary order. Now, let us prove Eq. (\ref{EQ59}). To this end we
compute how the r.h.s. of this equation acts on the state vector. We will use the
following factorization of the symmetrization operator
\be
S_N =\frac{(N-s)!}{N!}\sum_{\l_1<\ldots<\l_s}\prod_{\l_k}\xi^{\l_k-1}
\sum_{P_s}\xi(P_s) P_s S^{[s+1\ldots N]}_{N-s},
\en{EQ60}
where a permutation of $N$ elements is factorized into a product of the
permutations $P_s$ of the first $s$ and the last $N-s$ elements (summed up to the
symmetrization operator $ S^{[s+1\ldots N]}_{N-s}$) and a permutation between these
two subsets (given by the summation over the ordered set of indices
$\l_1,\ldots,\l_s$). By using Eq. (\ref{EQ60}) we obtain (the change of order in
the annihilation operator indices is made for convenience of notations)
\[
a_{\beta_1}\cdot\ldots\cdot a_{\beta_s}\vr{k_1,\ldots,k_N}
=[N(N-1)\cdot\ldots\cdot(N-s+1)]^{-\frac12}
\]
\[
\times\sum_{\l_1<\ldots<\l_s}\prod_{\l_k}\xi^{\l_k-1}\sum_{P}\xi(P)
\langle\beta_1|k_{P(\l_1)}\rangle\cdot\ldots\cdot\langle\beta_s|k_{P(\l_s)}\rangle
\underbrace{\vr{k_1,\ldots,k_N}}_{\vee k_{\l_1}\ldots \vee  k_{\l_s}}.
\]
Now we must calculate how the creation operators of Eq. (\ref{EQ59}) act on this
result. Applying consecutively  the definition  of the creation operator, Eq.
(\ref{EQ13}),  we get
\[
a^\d_{\alpha_s}\cdot\ldots\cdot a^\d_{\alpha_1}a_{\beta_1}\cdot\ldots\cdot
a_{\beta_s}\vr{k_1,\ldots,k_N}
\]
\[
=\sum_{\l_1<\ldots<\l_s}\prod_{\l_k}\xi^{\l_k-1}\sum_{P}\xi(P)
\langle\beta_1|k_{P(\l_1)}\rangle\cdot\ldots\cdot\langle\beta_s|k_{P(\l_s)}\rangle
\]
\[
\times S_N\{\vr{\alpha^{(1)}_s}\cdot\ldots\cdot\vr{\alpha^{(s)}_1}
\underbrace{\vr{k_1,\ldots,k_N}}_{\vee k_{\l_1}\ldots \vee k_{\l_s}}\}
\]
\[
=\sum_{\l_1<\ldots<\l_s}\sum_{P}\langle\beta_{P^{-1}(1)}|k_{\l_1}\rangle\cdot\ldots\cdot\langle\beta_{P^{-1}(s)}|k_{\l_s}\rangle
\underbrace{\vr{k_1,\ldots,k_N}}_{k_{\l_1}\to\alpha_{P^{-1}(1)}\ldots
k_{\l_s}\to\alpha_{P^{-1}(s)}},
\]
where we have transferred the permutation $P$ to the indices of $\beta$'s and used
the property $PS_N = \xi(P)S_N$ to transfer the signature of this permutation to
action  of this permutation  on $\alpha$-vectors. The final step is to note that
the last line is exactly the action of the l.h.s. of Eq. (\ref{EQ59}) on the state
vector.

Now let us expand the $s$-particle operator in the product of the  single-particle
basis states. We have
\[
A_s = \sum_{\alpha_1, \ldots, \alpha_s}\sum_{\beta_1,\ldots, \beta_s}
\sum_{j_1<\ldots<j_s}
\vr{\alpha^{(j_1)}_1}\cdot\ldots\cdot\vr{\alpha^{(j_s)}_s}A_s(\alpha_1,\ldots,\alpha_s|\beta_s,\ldots,\beta_1)
\]
\be
\times \vl{\beta^{(j_s)}_s}\cdot\ldots\cdot\vl{\beta^{(j_1)}_1}
\en{EQ61}
where the operator is represented by a symmetric function in the simultaneous
permutation of both $\alpha$'s and $\beta$'s:
\[
A_s(\alpha_1,\ldots,\alpha_s|\beta_s,\ldots,\beta_1) \equiv
\vl{\alpha^{(1)}_1}\cdot\ldots\cdot\vl{\alpha^{(s)}_s} A^{(1,...,s)}
\vr{\beta^{(s)}_s}\cdot\ldots\cdot\vr{\beta^{(1)}_1}.
\]
Therefore, adding a summation over the permutations of $s$ elements and  using Eq.
(\ref{EQ59}) we obtain
\be
A_s = \frac{1}{s!}\sum_{\alpha_1, \ldots, \alpha_s}\sum_{\beta_1,\ldots, \beta_s}
A_s(\alpha_1,\ldots,\alpha_s|\beta_s,\ldots,\beta_1)
a^\d_{\alpha_1}\cdot\ldots\cdot a^\d_{\alpha_s} a_{\beta_s}\cdot\ldots\cdot
a_{\beta_1}.
\en{EQ62}

One commentary on the identity (\ref{EQ59}) is in order. On the l.h.s. inside the summation we have an 
operator, the product of   ket- and bra-vectors,  which is not symmetrized in the indices $\alpha$ or $\beta$, whereas  on
the r.h.s. the operator is symmetrized, since by the commutation relations, we have
$P a^\d_{\alpha_1}\cdot\ldots\cdot a^\d_{\alpha_s} =
\xi(P)a^\d_{\alpha_P^{-1}(1)}\cdot\ldots\cdot a^\d_{\alpha_P^{-1}(s)}$. In fact,
due to this property of the r.h.s., one can apply the symmetrization to the vectors
on the l.h.s. and obtain the following identity
\be
\sum_{j_1<\ldots<j_s}\vr{\alpha_{1},\ldots,\alpha_{s}}
\vl{\beta_{s},\ldots,\beta_{1}} = \frac{1}{s!}a^\d_{\alpha_1}\cdot\ldots\cdot
a^\d_{\alpha_s} a_{\beta_s}\cdot\ldots\cdot a_{\beta_1},
\en{EQ63}
where the vectors on the l.h.s. are composed as  symmetrized products of the
vectors from the single-particle spaces with the indices $j_1,\ldots, j_s$.

\subsection{Examples}
\label{subI5B}

Consider the (one-dimensional) position and momentum operators of a system of $N$
indistinguishable particles. The observables formed of these two single-particle
operators must be symmetric with respect to permutation of the particles, thus the
only form they have is that of a sum:
\be
X = \frac{1}{N}\sum_{j=1}^Nx^{(j)},\quad  P = -i\hbar\sum_{j=1}^N \frac{\p}{\p
x^{(j)}},
\en{EQ64}
where we have used that the total momentum is a sum of the momenta of all the
particles and defined a the coordinate to be canonically conjugate to the momentum
operator, i.e. $[P,X] = -i\hbar$. Let us find the ``second-quantized'' form of this
operators. Define the creation and annihilation operators of the co-ordinate state
$|x\rangle$ as $\psi^\d(x)$ and $\psi(x)$. Then, the form of the  one-particle
operators $X$ and $P$ is given by  Eq. (\ref{EQ56}) (the summation is replaced by
the integration):
\be
X = \frac{1}{N}\int\rd x\,x\psi^\d(x)\psi(x),\quad  P = \int\rd
x\,\psi^\d(x)\left(-i\hbar\frac{\p}{\p x}\right)\psi(x),
\en{EQ65}
where, to arrive at the second formula, we have used that \mbox{$\langle
x^\prime|\frac{\p}{\p x}|x\rangle = -\frac{\p}{\p x}\delta(x^\prime-x)$} and used
the integration by parts. To clarify what physical entity corresponds to the
position observable $X$  notice that the number of particles is given by the
following formula
\be
N = \int\rd x\,\psi^\d(x)\psi(x).
\en{EQ66}
Hence $\rho(x) \equiv \psi^\d(x)\psi(x)$  is the operator of the density of
particles at $x$ (i.e. the average density is $\langle S|\rho(x)|S\rangle$, for a
state $|S\rangle$) and $\rho(x)/N$ gives  the probability density  to find a
particle at this point. Therefore, $X$  indeed describes the distribution of
position of the particles in the system.

Consider now a system of $N$ indistinguishable particles interacting by some
forces. We have in this case the Hamiltonian
\be
H = -\frac{\hbar^2}{2m}\sum_{j=1}^N \frac{\p^2}{(\p \br^{(j)})^2} + \sum_{1\le
i<j\le N}U(|\br^{(i)}-\br^{(j)}|).
\en{EQ67}
Using Eq. (\ref{EQ62}) we get for the interaction potential (the integration
$\int\rd^3 \br (...)$ is over the volume)
\[
\sum_{1\le i<j\le N}U(|\br^{(i)}-\br^{(j)}|) =
\frac12\int\rd^3\br\int\rd^3\br^\prime
U(|\br-\br^\prime|)\psi^\d(\br)\psi^\d(\br^\prime)\psi(\br^\prime)\psi(\br),
\]
which has a clear physical meaning, since the integrand is almost $\frac12
U(|\br-\br^\prime|)\rho(\br)\rho(\br^\prime)$ (except for the auto-interaction term
$-\frac12
U(|\br-\br^\prime|)\delta(\br-\br^\prime)\psi^\d(\br^\prime)\psi(\br^\prime)$),
which describes the interaction of a concentration of particles at $\br$ with that
at $\br^\prime$ by the potential $U(|\br-\br^\prime|)$ with the account of the
indistinguishability given by the coefficient $\frac12$. Hence, the Hamiltonian
reads
\[
H = \int\rd^3
\br\,\psi^\d(\br)\left(-\frac{\hbar^2}{2m}\nabla^2\right)\psi(\br)\qquad\qquad
\]
\be
+\frac12\int\rd^3\br\int\rd^3\br^\prime
U(|\br-\br^\prime|)\psi^\d(\br)\psi^\d(\br^\prime)\psi(\br^\prime)\psi(\br).
\en{EQ68}

%%%%%%%%%%%%%%%%%%%%%%%%%%%%%%%%%%%%%%%%%%%%%%%%%%%%%%%%%%%%%%%%%%%%%%%%%%%%%%%%%%%%%%

\section{Evolution of   operators in the Heisenberg picture}
\label{secI6}
Before we present the evolution of the creation and annihilation operators, let us
show the following useful identity about the commutator in the case of fermions. We
have
\[
[A,B_1B_2\cdot\ldots\cdot B_{2n}] = \{A,B_1\} B_2\cdot\ldots\cdot B_{2n} -
B_1\{A,B_2\} B_3\cdot\ldots\cdot B_{2n}
\]
\[
+\ldots +(-1)^{2n-1} B_1\cdot\ldots\cdot B_{2n-1}\{A,B_{2n}\},
\]
where $\{A,B\} = AB+BA$ is the anti-commutator. This identity  can be verified by
induction. Since in the number of  particle preserving Hamiltonian the number of
creation and annihilation fermion operators is the same, the above identity allows
one to expand the commutator $[H,A]$ using the anti-commutators of $A$ with the
creation and annihilation operators.  We will use then the unified notation
$[A,B]_\xi$ which denotes the commutator in the case of bosons and anti-commutator
in the case of fermions, with this notation and the replacement of $(-1)$ by $\xi$
the above identity  becomes  universally valid for bosons and fermions:
\[
[A,B_1B_2\cdot\ldots\cdot B_{2n}] = [A,B_1]_\xi B_2\cdot\ldots\cdot B_{2n} +\xi
B_1[A,B_2]_\xi B_3\cdot\ldots\cdot B_{2n}
\]
\be
+\ldots +\xi^{2n-1} B_1\cdot\ldots\cdot B_{2n-1}[A,B_{2n}]_\xi,
\en{EQ69}

In the Heisenberg picture, the evolution is transferred to the observables, i.e. we
have the operator equation
\be
i\hbar\frac{\p}{\p t}A = [A,H],
\en{EQ70}
where $A$ is any observable of the quantum system and $H$ is its Hamiltonian. Let
us use Eq. (\ref{EQ70}) for the annihilation operator $\psi(\br)$ for the system of
indistinguishable particles with the Hamiltonian given by Eq. (\ref{EQ68}). Using
Eq. (\ref{EQ69}) to compute the commutators:
\[
[\psi(\br),\psi^\d(\br_1)\left(-\frac{\hbar^2}{2m}\frac{\p^2}{\p
\br_1^2}\right)\psi(\br_1)] = [\psi(\br),\psi^\d(\br_1)]_\xi
\left(-\frac{\hbar^2}{2m}\frac{\p^2}{\p \br_1^2}\right)\psi(\br_1)
\]
\[
+ \xi\psi^\d(\br_1)[\psi(\br),\left(-\frac{\hbar^2}{2m}\frac{\p^2}{\p
\br_1^2}\right)\psi(\br_1)]_\xi =-\delta(\br-\br_1)\frac{\hbar^2}{2m}\frac{\p^2}{\p
\br_1^2}\psi(\br_1),
\]
where we have used the linearity of the commutator (to apply the linear operator
$\frac{\p^2}{\p \br_1^2}$ after the evaluation of the commutator itself) and the
commutation rules of the creation and annihilation operators, and
\[
[\psi(\br),\psi^\d(\br_1)\psi^\d(\br_2)\psi(\br_2)\psi(\br_1)] =
\delta(\br-\br_1)\psi^\d(\br_2)\psi(\br_2)\psi(\br_1)+\xi\delta(\br-\br_2)\psi^\d(\br_1)\psi(\br_2)\psi(\br_1)
\]
\[
=\left[\delta(\br-\br_1)\psi^\d(\br_2)\psi(\br_2) +
\delta(\br-\br_1)\psi^\d(\br_1)\psi(\br_1)\right]\psi(\br),
\]
where we have used the commutation rules of the creation nd annihilation operators
(e.g.  the two annihilation operators in the second term on the r.h.s. we replaced
in the second line). Using these results we obtain from Eqs. (\ref{EQ70}) and
(\ref{EQ68}) the following evolution equation for the  annihilation operator
\be
i\hbar\frac{\p}{\p t}\psi(\br) =-\frac{\hbar^2}{2m}\frac{\p^2}{\p \br^2}\psi(\br)+
\left(\int\rd^3\br^\prime
U(|\br-\br^\prime|)\psi^\d(\br^\prime)\psi(\br^\prime)\right)\psi(\br).
\en{EQ71}
Note that the second term defines a common potential for the particles through the
particle density operator $\rho(\br) =\psi^\d(\br)\psi(\br)$, i.e.
\be
\mathcal{U}(\br)\equiv \int\rd^3\br^\prime \rho(\br^\prime)U(|\br-\br^\prime|),
\en{EQ72}
so that the evolution equation may be cast in the form resembling that of the usual
Schrodinger equation
\be
i\hbar\frac{\p}{\p t}\psi(\br) =-\frac{\hbar^2}{2m}\frac{\p^2}{\p
\br^2}\psi(\br)+\mathcal{U}(\br)\psi(\br).
\en{EQ73}
This is one of the reasons the method is sometimes called the second quantization
(one can say that we have performed the ``second quantization'' of  the Schrodinger
equation to arrive at Eq. (\ref{EQ73})).

%%%%%%%%%%%%%%%%%%%%%%%%%%%%%%%%%%%%%%%%%%%%%%%%%%%%%%%%%%%%%%%%%%%%%%%%%%%%%%%%%%%%%%%%%%%%

\section{Statistical  operators of indistinguishable particles}
\label{secI7}

The method of statistical operators was developed by N.N.~Bogoliubov \cite{BgBook}.
Suppose that we describe a system of $N\gg1$ indistinguishable particles and  are
interested in the $s$-particle operators, with $s$ being  much less then $N$
(usually, $s=1$ or $2$). Being interested in the few-particle observables, instead
of using the full $N$-particle state of the system (or the full density matrix) one
can use the so-called statistical operators of particle complexes (or the
statistical operators of Bogoliubov)  which fully describe the properties of such
observables.  For simplicity we assume that the $N$-particle system is in the pure
state $|\Psi_N\rangle$ (using the density matrix is a straightforward
generalization). Expanding it in the physical basis, i.e.
\be
|\Psi_N\rangle =  \sum_{\{n_l\}}\Phi_{\{n_l\}}|\{n_l\}\rangle,
\en{EQ74}
where the set $\{n_l\}$ is  the corresponding set of occupation numbers of the
basis state. The $s$-particle statistical operator is given by the partial trace of
the symmetrized (i.e. physical) state of Eq. (\ref{EQ74}) over the $N\!-\!s$
single-particles Hilbert spaces, i.e.
\[
\mathcal{R}_s \equiv\Tr_{s+1,\ldots,N}\{\vr{\Psi_N}\vl{\Psi_N}\}
\]
\be
=\sum_{k_{s+1}}\cdot\ldots\cdot\sum_{k_N}
\left[\vl{k^{(s+1)}_{s+1}}\cdot\ldots\cdot\vl{k^{(N)}_N}\right]\vr{\Psi_N}\vl{\Psi_N}
\left[\vr{k^{(s+1)}_{s+1}}\cdot\ldots\cdot\vr{k^{(N)}_N}\right],
\en{EQ75}
where the vectors $\vr{k_j}$ are the  orthogonal  basis states in the
single-particle Hilbert space. Due to the property of the symmetrization operators
$S_NS_{N-s} = S_N$ one can use in Eq. (\ref{EQ75}) the (unnormalized) symmetrized
basis vectors instead. Moreover, one can use the identity given by Eq.
(\ref{EQ44}), but now for the $N-s$ particles, to arrive at the following more
simple expression of the statistical operator (with the drawback, however,  that
the notations do not indicate over which single-particle spaces the inner product
is taken)
\be
\mathcal{R}_s =
\sum_{\sum\tilde{n}_l=N-s}\langle\{\tilde{n}_l\}|\Psi_N\rangle\langle
\Psi_N|\{\tilde{n}_l\}\rangle,
\en{EQ76}
where $|\{\tilde{n}_l\}\rangle$ is the normalized symmetrized vector made of
$\vr{k^{(s+1)}_{s+1}}\cdot\ldots\cdot\vr{k^{(N)}_N}$.

In the special case, when the state of the system is one of the basis states,
$\vr{\Psi_N} = |\{n_l\}\rangle$, one can derive an insightful  explicit expression
for the $s$-particle operator. In this case, the $s$-particle operator of Eq.
(\ref{EQ75}) can be cast as follows
\be
\mathcal{R}_s = \sum_{\sum\limits_l m_l=s}
\frac{\prod\limits_lC^{m_l}_{n_l}}{C^s_N}\vr{\{m_l\}}\vl{\{m_l\}},
\en{EQ77}
where $C^k_n = \frac{n!}{(n-k)!k!}$ is the usual combinatorial coefficient  and the
vector $\vr{\{m_l\}}$ is the short-hand notation for  a state of   $s$ particles
with the occupation numbers $m_l\le n_l$, with $n_l$ being the occupation numbers
of the system state. This result is valid for both bosons and fermions, in the
latter case, however, one can drop the combinatorial coefficient in the numerator
since it is equal to 1 (and all the occupation numbers $m_l$ are also equal to
either 0 or 1). Note that the physical meaning of the scalar factor in Eq.
(\ref{EQ77}) is the probability of finding $s$ particles exactly  in the state
$|\{m_l\}\rangle\langle \{m_l\}|$, where the particles/states are drawn from the
state of the system $|\{n_l\}\rangle$. Indeed, let us show that the sum of these
coefficients (probabilities) is exactly 1. This follows from the following identity
\[
(x+y)^{N} = \prod\limits_l(x+y)^{n_l} =
\prod\limits_l\left(\sum_{m_l=0}^{n_l}C^{m_l}_{n_l}x^{m_l}y^{n_l-m_l} \right)
\]
\[
=\sum_{\{m_l\}}\prod\limits_lC^{m_l}_{n_l}x^{\sum\limits_lm_l}y^{\sum\limits_l(n_l-m_l)}.
\]
Taking the partial derivative, i.e. applying the operator
$\frac{1}{s!}\frac{1}{(N-s)!}\frac{\p^s}{\p x^s}\frac{\p^{N-s}}{\p y^{N-s}}$ at
$(x=0, y=0)$ we get
\[
\frac{N!}{s!(N-s)!} = \sum_{\sum\limits_l m_l=s} \prod\limits_lC^{m_l}_{n_l},
\]
where $\sum_l n_l = N$.

To arrive at the  explicit expression for the $s$-particle statistical operator
(\ref{EQ77}), we will use the property of the symmetrization operator given by Eq.
(\ref{EQ60}). We have then
\[
\mathcal{R}_s =
\frac{N!}{\prod\limits_ln_l!}\left(\frac{(N-s)!}{N!}\right)^2\sum_{l_1<\dots<l_s}\prod_{l_k}\xi^{l_k-1}
\sum_{\l_1<\ldots<\l_s}\prod_{\l_k}\xi^{\l_k-1}
\sum_{P_s}\xi(P_s)\sum_{P_s^\prime}\xi(P_s^\prime)
\]
\be
\times\frac{\prod\limits_l(n_l-m_l)!}{(N-s)!}\prod\limits_l\delta_{m_l,m^\prime_l}
P_s\{\vr{f_{j_1}}\cdot\ldots\cdot\vr{f_{j_s}}\}P^\prime_s\{\vl{g_{j_1}}\cdot\ldots\cdot\vl{g_{j_s}}\}
\en{EQ78}
where we have used Eq. (\ref{EQ60}) to rewrite each basis vector
$|f_1,\ldots,f_N\rangle$ from the expansion of the system state $\vr{\Psi_N}$ in
the following form
\[
|f_1,\ldots,f_N\rangle
=\frac{(N-s)!}{N!}\sum_{l_1<\dots<l_s}\prod_{l_k}\xi^{l_k-1}\sum_{P_s}\xi(P_s)P_s\{\vr{f^{(1)}_{l_1}}\cdot\ldots\cdot\vr{f^{(s)}_{l_s}}\}
\]
\[
\times S_{N-s}\{\vr{f^{(s+1)}_{l_{s+1}}}\cdot\ldots\cdot\vr{f^{(N)}_{l_N}}\}
\]
and introduced  the  normalized vectors $|\{m_l\}\rangle$ and $|\{n_l-m_l\}\rangle$
\[
|\{m_l\}\rangle =\left(\frac{s!}{\prod\limits_lm_l!}\right)^\frac12
S_s\{\vr{f_{l_1}}\cdot\ldots\cdot\vr{f_{l_s}}\},
\]
\[
|\{n_l-m_l\}\rangle =\left(\frac{(N-s)!}{\prod\limits_l(n_l-m_l)!}\right)^\frac12
S_{N-s}\{\vr{f_{l_{s+1}}}\cdot\ldots\cdot\vr{f_{l_N}}\},
\]
corresponding to the partition of a set $(f_1,\ldots,f_N)$ into
$(f_{l_1},\ldots,f_{l_s})$ and $(f_{l_{s+1}},\dots,f_{l_N})$. Taking the trace as
in Eq. (\ref{EQ76}) is equivalent to setting $\tilde{n}_l = n_l-m_l$ and
multiplying by the inverse of the corresponding normalization factor
${\prod\limits_l(n_l-m_l)!}/{(N-s)!}$. Now using in Eq. (\ref{EQ78}) the following
summation identity
\[
\sum_{l_1<\ldots<l_s}(\ldots) =
\sum_{\{m_l\}}\prod\limits_l\frac{n_l!}{m_l!(n_l-m_l)!}(\ldots),
\]
and the fact that in the case of fermions the partitions of $f$'s and $g$'s  should
be the same (due to the delta-symbol in EQ. (\ref{EQ78}),   in this case the
signature $\xi^{l_k-1}$ appears twice and cancels out) and in the case of bosons
$\xi=1$, by rewriting the vectors in the normalized form, we get Eq. (\ref{EQ77}).

It is now quite clear that in the general case, when the $N$-particle state is a
linear combination of the Fock-basis states Eq. (\ref{EQ74}). In this case,  the
$s$-particle statistical operator can be cast in the following form
\[
\mathcal{R}_s =
\sum_{\{n_l\}}\sum_{\{n^\prime_l\}}\Phi^*(\{n^\prime_l\})\Phi(\{n_l\})\sum_{\sum\limits_lm_l=s}\;\sum_{\sum\limits_lm^\prime_l=s}
\]
\[
\times\prod\limits_l\left(\delta_{n_l-m_l,n^\prime_l-m^\prime_l}\frac{\sqrt{C^{m^{}_l}_{n_l}C^{m^\prime_l}_{n^\prime_l}}}{C^s_N}\right)
\vr{\{m_l\}}\vl{\{m^\prime_l\}},
\]
or by setting $p_l \equiv n_l-m_l=n^\prime_l-m^\prime_l$ and using that
$C^{m^{}_l}_{n_l} = C^{p^{}_l}_{n_l}$ we get
\be
\mathcal{R}_s =
\sum_{\{n_l\}}\sum_{\{n^\prime_l\}}\Phi^*(\{n_l\})\Phi(\{n^\prime_l\})\sum_{\sum\limits_lp_l=N-s}
\frac{\prod\limits_l\sqrt{C^{p_l}_{n_l}C^{p_l}_{n^\prime_l}}}{C^s_N}
\vr{\{n_l-p_l\}}\vl{\{n^\prime_l-p_l\}},
\en{EQ80}
where if  $p_l> \mathrm{min}(n_l,n^\prime_l)$ the corresponding term is simply
zero.

There is yet another general representation for the $s$-particle statistical
operator, which follows directly from Eq. (\ref{EQ80}). Indeed, each  vector in the
product in Eq. (\ref{EQ80}) can be represented by action of the annihilation
operator to the corresponding  basis vector, i.e.,
\[
\prod\limits_l\sqrt{C^{p_l}_{n_l}}\vr{\{n_l-p_l\}}
=\prod\limits_l\frac{a^{p_l}_l}{\sqrt{p_l!}}\vr{\{n_l\}},
\]
(in the case of fermions the product of the annihilation operators is ordered one,
however, since there is a bra-vector in Eq. (\ref{EQ80}) the order does not matter)
thus we have Eq. (\ref{EQ80}) rewritten in the form
\[
\mathcal{R}_s =
\frac{1}{C^s_N}\sum_{\{n_l\}}\sum_{\{n^\prime_l\}}\Phi^*(\{n^\prime_l\})\Phi(\{n_l\})\sum_{\sum\limits_lp_l=N-s}
\prod\limits_l\frac{a^{p_l}_l}{\sqrt{p_l!}}
\vr{\{n_l\}}\vl{\{n^\prime_l\}}\frac{(a^\d_l)^{p_l}}{\sqrt{p_l!}},
\]
where the order of the creation is inverse to the order of the annihilation
operators. The general result is thus given by the following simple formula
\be
\mathcal{R}_s = \frac{1}{C^s_N}\sum_{\sum\limits_lp_l=N-s} \prod\limits_l
\frac{a^{p_l}_l\rho_N(a^\d_l)^{p_l}}{p_l!},
\en{EQ80A}
were $\rho_N$ is the general $N$-particle density matrix (in our case $\rho_N =
\vr{\Psi_N}\vl{\Psi_N}$) and, in the case of fermions,  the order of the operators
on the left and on the right of it is inverse to each other.

%%%%%%%%%%%%%%%%%%%%%%%%%%%%%%%%%%%%%%%%%%%%%%%%%%%%%%%%%%%%%%%%%%%%%%%%%%%%%%%%%%

\subsection{The averages of the $s$-particle operators }
\label{subI7A}

The average of the $s$-particle operator is defined as the trace of it with the
state vector (or density matrix, in general). We have
\be
\langle A_s\rangle \equiv \Tr\{A_s|\Psi_N\rangle\langle \Psi_N|\}.
\en{EQ81}
We have using the expansion of the operator unit in the physical space,
Eq.~(\ref{EQ45}),
\[
\langle A_s\rangle  = \sum_{k_1}\cdot\ldots\cdot\sum_{k_N}\vl{k_1,\ldots,k_N}A_s
|\Psi_N\rangle\langle \Psi_N\vr{k_1,\ldots,k_N}
\]
\be
=C^s_N\sum_{k_1}\cdot\ldots\cdot\sum_{k_N}\vl{k_1,\ldots,k_N}A^{(1,\ldots,s)}
|\Psi_N\rangle\langle \Psi_N\vr{k_1,\ldots,k_N},
\en{EQ82}
where we have used the symmetry properties of the physical vectors and the symmetry
property of the $s$-particle operator, i.e. $A^{(P^{-1}(1),\ldots,P^{-1}(s))} =
P^\d A^{(1,\ldots,s)}P$ and   the number of all partitions $C^s_N$ of the set
$(1,\ldots,N)$ into $(j_1,\ldots,j_s)$ and its complementary subset. Taking into
account the definition of the $s$-particle statistical operator, Eq. (\ref{EQ76}),
we get finally
\be
\langle A_s\rangle = C^s_N\Tr\{ A^{(1,\ldots,s)} \mathcal{R}_s\}.
\en{EQ83}
Indeed, we have to verify that the partition of the summation in Eq. (\ref{EQ82})
into two subsets, using Eq. (\ref{EQ60}),
\be
\vl{k_1,\ldots,k_N} =
\frac{1}{C^s_N}\sum_{j_1<\ldots<j_s}\prod\limits_l\xi^{j_l-1}\vl{k_{j_1},\ldots,k_{j_s}}\vl{k_{j_{s+1}},\ldots,k_{j_{N}}}
\en{EQ84}
leads to Eq. (\ref{EQ83}). But this is evident from the following considerations.
First, by the symmetry with respect to the permutations of $s$ elements in
$A^{(1,\ldots,s)}$ and the symmetry properties of the physical vectors we get
\[
\vl{k_{j_1},\ldots,k_{j_s}}\vl{k_{j_{s+1}},\ldots,k_{j_{N}}}A^{(1,\ldots,s)}\vr{\Psi_N}=
\bigl[ \vl{k_{j_1}}\cdot\ldots\cdot\vl{k_{j_N}}\bigr]A^{(1,\ldots,s)}\vr{\Psi_N},
\]
\[
\langle{\Psi_N}\vr{k_{j_1},\ldots,k_{j_N}} = \langle{\Psi_N}|\bigl[
\vr{k_{j_1}}\cdot\ldots\cdot\vr{k_{j_N}}\bigr].
\]
Second, by noticing that the summation over the ordered set $l_1<\ldots<l_s$ gives
exactly the $C^s_N$ equal terms, which cancels the coefficient $1/C^s_N$ in Eq.
(\ref{EQ84})  (for the fermion case, the signature cancels out too since the
single-particle states $\vr{k_j}$ used to build the $N$-particle state are all
different). We arrive then at Eq. (\ref{EQ83}).

Consider now the particular choice of the $s$-particle operator given through  the
$s$-partition of the co-ordinate basis  vectors $\vr{x_1,\ldots,x_N}$ (we consider
the one-dimensional case for the simplicity of formulae below) as follows
\be
X_s \equiv \sum_{j_1<\ldots<j_s}S_s\{\vr{x^{(j_1)}_{1},\ldots,x^{(j_s)}_{s}}\}
S_s\{\vl{\tilde{x}^{(j_1)}_{1},\ldots,\tilde{x}^{(j_s)}_{s}}\}.
\en{EQ85}
Obviously, $X_s$ is a symmetric operator, hence an observable in the physical
space. Then by Eq. (\ref{EQ63}) we get a simpler representation for $X_s$:
\be
X_s =
\frac{1}{s!}\psi^\d(x_1)\cdot\ldots\cdot\psi^\d(x_s)\psi(\tilde{x}_s)\cdot\ldots\cdot\psi(\tilde{x}_1).
\en{EQ86}
On the other hand, by using Eqs. (\ref{EQ83}) with $A_s = X_s$ and Eq. (\ref{EQ85})
we obtain the identity
\[
\langle X_s\rangle
=C^s_N\vl{\tilde{x}_1,\ldots,\tilde{x}_s}\mathcal{R}_s\vr{x_1,\ldots,x_s}
\]
\[
=C^s_N\mathcal{R}_s(\tilde{x}_1,\ldots,\tilde{x}_s|x_1,\ldots,x_s).
\]
Therefore, we have arrived at the important  identity for the $s$-particle
statistical operator in the co-ordinate representation:
\be
\mathcal{R}_s(\tilde{x}_1,\ldots,\tilde{x}_s|x_1,\ldots,x_s) =
\frac{(N-s)!}{N!}\langle\psi^\d(x_1)\cdot\ldots\cdot\psi^\d(x_s)\psi(\tilde{x}_s)\cdot\ldots\cdot\psi(\tilde{x}_1)\rangle.
\en{EQ87}

Finally, by using the co-ordinate version of the general representation of a
$s$-particle operator, Eq. (\ref{EQ62}), one can express the average of the
$s$-particle observable also as follows
\[
\langle A_s\rangle = \frac{1}{s!}\int\rd x_1\cdot\ldots\cdot\int\rd x_s\int\rd
\tilde{x}_1\cdot\ldots\cdot\int\rd\tilde{x}_sA_s(\tilde{x}_1,\ldots,\tilde{x}_s|x_1,\ldots,x_s)
\]
\be
\times\langle\psi^\d(x_1)\cdot\ldots\cdot\psi^\d(x_s)\psi(\tilde{x}_s)\cdot\ldots\cdot\psi(\tilde{x}_1)\rangle.
\en{EQ88}
By comparing with Eq. (\ref{EQ83}) we again obtain the result of Eq. (\ref{EQ87}).

When the total number of particles in the system is not fixed   it is more
convenient to work with the non-normalized statistical operators. First of all, the
notion of the statistical operator $\mathcal{R}_s$ can be easily generalized to
such systems. Indeed,  one observes  observing that $\mathcal{R}_s$ is linear in
the state of the system, i.e. if the system is described by the density matrix
$\rho = \sum_N p_N \rho_N$, $0\le p_N\le1$ then $\mathcal{R}_s = \sum_N p_N
\mathcal{R}^{(N)}_s$, where each $\mathcal{R}^{(N)}$ is defined for $\rho_N$ by Eq.
(\ref{EQ75}) with $|\Psi_N\rangle\langle \Psi_N|$ substituted by $\rho_N$.

Note that, for instance,  the diagonal part of the  $1$-particle statistical
operator $\mathcal{R}_1$ in the co-ordinate representation is proportional to the
density $\mathcal{R}_1(\br|\br) = N^{-1}\langle \psi^\d(\br)\psi(\br)\rangle$, thus
it is convenient to introduce the operator  $\sigma_1$ by $\sigma_1(\br^\prime|\br)
= N\mathcal{R}_1(\br|\br^\prime) = \langle \psi^\d(\br)\psi(\br^\prime)\rangle$
which does not involve the total number of particles directly and is applicable to
the systems with variable number of particles. The $s$-particle statistical
operator $\sigma_s$ can be thus defined through the $\mathcal{R}_s$  by the
following formula
\be
\sigma_s = \sum_N p_N \frac{N!}{(N-s)!}\mathcal{R}^{(N)}_s,
\en{sigmas}
or, using Eq. (\ref{EQ87}), we get  in the coordinate representation (in the three
dimensions now)
\be
\sigma_s(\tilde{\br}_1,\ldots,\tilde{\br}_s|\br_1,\ldots,\br_s) =
\langle\psi^\d(\br_1)\cdot\ldots\cdot\psi^\d(\br_s)\psi(\tilde{\br}_s)\cdot\ldots\cdot\psi(\tilde{\br}_1)\rangle.
\en{sigm}

%%%%%%%%%%%%%%%%%%%%%%%%%%%%%%%%%%%%%%%%%%%%%%%%%%%%%%%%%%%%%%%%%%%%%%%%%%%%%%%%%%%%%%%%
\subsection{The general structure of the  one-particle statistical operator}
\label{subI7B}

 Consider a system of $N$ indistinguishable particles in a pure state
(\ref{EQ74}), i.e.
\[
|\Psi_N\rangle =  \sum_{\{n_l\}}\Phi_{\{n_l\}}|\{n_l\}\rangle.
\]
The goal is to find the general expression for the one-particle statistical
operator $\mathcal{R}_1$. We will use the expression (\ref{EQ80A}), i.e.
\[
\mathcal{R}_1 = \frac{1}{N}\sum_{\sum\limits_lm_l=N-1} \prod\limits_l
\frac{a^{m_l}_l\vr{\Psi_N}\vl{\Psi_N}(a^\d_l)^{m_l}}{m_l!}.
\]
We have here
\[
\prod\limits_{\{l,{\sum\limits_lm_l=N-1}\}}
\frac{a^{m_l}_l\vr{\{n_l\}}\vl{\{\tilde{n}_l\}}(a^\d_l)^{m_l}}{m_l!}
=\sum_{l_1,l_2}\sqrt{n_{l_1}\tilde{n}_{l_2}}\vr{f_{l_1}}\vl{f_{l_2}}
\]
where  the occupation number  $n_l$  corresponds to the state $\vr{f_{l}}$ (in the
case of bosons $n_{l_s}\ge1$ and for fermions $n_{l_s}=1$). Since we also
``remove'' equal number of particles from each  $l$th-state in the bra and ket
vectors, there are just two cases: i) there is such $l=\ell$ that $m_\ell =
n_\ell-1$, while for $l\ne \ell$ $m_l = n_l$, and the same is true for
$\tilde{n}_l$ and ii) there are two such $l_1$ and $l_2$ that $m_{l_1} =
\tilde{n}_{l_1}= n_{l_1}-1$ and $m_{l_2} = \tilde{n}_{l_2}-1= n_{l_2}$. Depending
on the $N$-particle state, we have a number of occurrences of both these cases in
the one-particle operator $\mathcal{R}_1$. Therefore, using the formula
(\ref{EQ80A}) we obtain the following general structure of the one-particle
operator $\mathcal{R}_1$:
\be
\mathcal{R}_1 = \sum_l\sum_{n=1}^N A^{(1)}_l({n})\frac{n}{N}\vr{f_l}\vl{f_l}+
\sum_{l_1\ne
l_2}\sum_{\;n_1\!,n_2=1}^NA^{(2)}_{l_1,l_2}({n_1,n_2})\frac{\sqrt{n_1n_2}}{N}\vr{f_{l_1}}\vl{f_{l_2}},
\en{EQ89}
with the coefficients given by the formulae
\[
A^{(1)}_l(n) = \sum_{\{n_j;j\ne l\}}|\Phi(n_1,\ldots,n_l=n,\ldots)|^2, \]
\[
A^{(2)}_{l_1,l_2}(n_1,n_2) =\sum_{\{n_j;j\ne
l_1,l_2\}}\Phi^*(n_1,\ldots,n_{l_1}\!=\!n_1\!-\!1,\ldots, n_{l_2}\!=\!n_2,\dots)
\]
\[
\times \Phi(n_1,\ldots,n_{l_1}\!=\!n_1,\ldots, n_{l_2}\!=\!n_2\!-\!1,\dots),
\]
where the summation is understood over all $n_j$ except for the selected indices.
The obvious property $(A^{(2)}_{l_1,l_2}(n_1,n_2))^* = A^{(2)}_{l_2,l_1}(n_2,n_1)$
guarantees that the the statistical operator is Hermitian, as it should be.
Moreover, it is easy to verify that it has the unit trace, since
\[
\sum_l\sum_{n=1}^NA^{(1)}_l(n)\frac{n}{N} =
\sum_{\{n_j\}}\left(\sum_l\frac{n_l}{N}\right)|\Phi(n_1,\ldots,n_l,\ldots)|^2 = 1.
\]

Finally, the non-normalized statistical operator $\sigma_1$ is given by the
following formula
\[
\sigma_1 = \sum_l\sum_{n}  A^{(1)}_l({n})n\vr{f_l}\vl{f_l}+ \sum_{l_1\ne
l_2}\sum_{\;n_1\!,n_2 }
A^{(2)}_{l_1,l_2}({n_1,n_2})\sqrt{n_1n_2}\vr{f_{l_1}}\vl{f_{l_2}}
\]
\be
=\sum_l\langle a^\d_l a_l\rangle\vr{f_l}\vl{f_l}+ \sum_{l_1\ne
l_2}\sum_{\;n_1\!,n_2 } \langle a^\d_{l_2} a_{l_1}\rangle\vr{f_{l_1}}\vl{f_{l_2}}.
\en{sigma1}
This representation clarifies the meaning of the  eigenvalues of the Hermitian
operator  $\sigma_1$: they are  the \textit{average occupation numbers} of the
corresponding eigenstates used as a Fock basis states, i.e. in the above
representation, for instance, we have $\langle n_l\rangle = \sum_{n}
A^{(1)}_l({n})n$ by the definition of $A^{(1)}_l(n)$.

%%%%%%%%%%%%%%%%%%%%%%%%%%%%%%%%%%%%%%%%%%%%%%%%%%%%%%%%%%%%%%%%%%%%%%%%%%%%%%%%%%%%%%%%%%%%%%%%%%%%%%
%%%%%%%%%%%%%%%%%%%%%%%        QUADRATIC HAMILTONIANS            %%%%%%%%%%%%%%%%%%%%%%%%%%%%%%%%%%%%
%%%%%%%%%%%%%%%%%%%%%%%%%%%%%%%%%%%%%%%%%%%%%%%%%%%%%%%%%%%%%%%%%%%%%%%%%%%%%%%%%%%%%%%%%%%%%%%%%%%%%%

\chapter{Quadratic Hamiltonian and the diagonalization }
\label{chII}

The properties of the  Hamiltonians quadratic in the creation and annihilation
operators are reviewed. The action of such a Hamiltonian is related to the
canonical transformation, which is given in the explicit form. In the case of
fermions the quadratic Hamiltonian can always be diagonalized by a unitary
transformation. In the case of bosons the problem of diagonalization of a quadratic
Hamiltonian is related to the old problem of stability of a solution to a
Hamiltonian system of the classical mechanics, for instance,  the notion of the
Krein index plays also a crucial role for  the diagonalization of the bosonic
Hamiltonian. A simple criterion, i.e. a necessary and sufficient condition, to the
latter problem is unknown, however, a sufficient condition for a quadratic bosonic
Hamiltonian to be diagonalizable is the positivity or negativity  of the related
classical Hessian. Also, the diagonalization of a quadratic boson Hamiltonian in
the case of a simplest (i.e. rank 2) zero mode is considered.

\newpage
\section{Diagionalization of the Hamiltonian quadratic in the fermion operators}
\label{secII1}

The general form of a quadratic Hamiltonian operator is    the following
\be
H = \sum_{\mu\nu}A_{\mu\nu}a^\d_\mu a_\nu + \frac12\sum_{\mu\nu}B_{\mu\nu}a_\mu
a_\nu-\frac12\sum_{\mu\nu}B^*_{\mu\nu}a^\d_\mu a^\d_\nu.
\en{EW1}
Here the indices run from 1 to $N$ and, due to the fermion anti-commutation
relations $\{a_\mu,a_\nu\} = 0$ and $\{a_\mu,a^\d_\nu\} = \delta_{\mu,\nu}$, the
matrices satisfy the properties $A^\d = A$ and $\wtilde{B} = -B$, where the tilde
denotes the matrix transposition, i.e. $\wtilde{B}_{\mu\nu} = B_{\nu\mu}$. Let us
define the operator-valued columns and rows, i.e.:
\be
\ba \equiv \left(\begin{array}{c}a_1\\ a_2\\ \vdots\\ a_N\end{array} \right),\quad
\wtilde{\ba}\equiv \left(\begin{array}{cccc}a_1,& a_2,& \ldots,& a_N\end{array}
\right),
\en{EW2}
and similar the column $\ba^\d$ and row  $\wtilde{\ba}^\d$.  Using the
anti-commutators relations and  that $\wtilde{A} = A^*$ we obtain
\[
\wtilde{\ba}^\d A\ba = - \wtilde{\ba}\wtilde{A}\ba^\d + \mathrm{Tr}(A)=-
\wtilde{\ba}A^*\ba^\d + \mathrm{Tr}(A).
\]
Hence, we can rewrite the Hamiltonian (\ref{EW1}) in the following matrix form
\be
H = \frac12\left(\begin{array}{cc}\wtilde{\ba}^\d,\wtilde{\ba}\end{array} \right)
\left(\begin{array}{cc}A & B^\d\\ B & -A^*\end{array} \right)\left(\begin{array}{c}
\ba\\ \;\ba^\d\end{array} \right) + \frac12\mathrm{Tr}(A).
\en{EW3}
Note the properties of  the traceless matrix $\mathcal{H}$ of  the fermionic
quadratic form (\ref{EW3}):
\be
\mathcal{H} \equiv \left(\begin{array}{cc}A &B^\d\\ B & -A^*\end{array} \right), \;
\mathcal{H}^\d = \mathcal{H}, \quad \tau\mathcal{H}\tau = -\mathcal{H}^*, \; \tau \equiv \left(\begin{array}{cc}0 & I_{N\times N} \\
I_{N\times N} & 0\end{array} \right),
\en{EW4}
i.e. $\tau$ is the block off-diagonal matrix, where each block has the dimension
$N\times N$. The matrix $\mathcal{H}$ could be a Hessian of a classical Hamiltonian
system (with a quadratic Hamiltonian) if it were not for the minus sign at the last
property (due to the anti-commutation relations between the fermion operators). We
however will call $\mathcal{H}$  the Hessian of a fermionic Hamiltonian. The
analogy with the classical Hamiltonian system is possible for the bosonic quadratic
Hamiltonian, considered in the next section.

We are interested in the possibility of diagonalizing of the Hamiltonian
(\ref{EW3}). Such a transformation, if it exists,  must preserve the
anti-commutation relations between the new fermion operators. A general linear
transformation between the creation and annihilation operators can be put in the
following form
\be
\left(\begin{array}{cc} \ba\\ \;\ba^\d\end{array} \right)= \left(\begin{array}{cc}U
& V^*\\ V & U^*\end{array} \right) \left(\begin{array}{cc} \bb\\
\;\bb^\d\end{array}\right)\equiv S\left(\begin{array}{cc} \bb\\
\;\bb^\d\end{array}\right),
\en{EW5}
where the matrices $U$ and $V$ are to be constrained  by the anti-commutations of
the involved operators. We have
\[
\{a_\mu,a_\nu\} =
\sum_{\alpha,\beta}\{U_{\mu\alpha}b_\alpha+V^*_{\mu\alpha}b^\d_\alpha,U_{\nu\beta}b_\beta+V^*_{\nu\beta}b^\d_\beta\}
=\sum_{\alpha}(U_{\mu\alpha}V^*_{\nu\alpha}+V^*_{\mu\alpha}U_{\nu\alpha})=0,
\]
\[
\{a_\mu,a^\d_\nu\} =
\sum_{\alpha,\beta}\{U_{\mu\alpha}b_\alpha+V^*_{\mu\alpha}b^\d_\alpha,U^*_{\nu\beta}b^\d_\beta+V_{\nu\beta}b_\beta\}
=\sum_{\alpha}(U_{\mu\alpha}U^*_{\nu\alpha}+V^*_{\mu\alpha}V_{\nu\alpha})=\delta_{\mu,\nu},
\]
i.e., in the matrix form, we have obtained the relations
\be
UU^\d+V^*\wtilde{V}=I,\quad UV^\d+V^*\wtilde{U}=0.
\en{EW6}
These conditions, in their turn, guarantee also the existence of the inverse
transformation to that of Eq. (\ref{EW5}), i.e.,
\be
\left(\begin{array}{cc} \bb\\ \;\bb^\d\end{array} \right)=
\left(\begin{array}{cc}U^\d &
V^\d\\ \wtilde{V} & \wtilde{U}\end{array} \right) \left(\begin{array}{cc} \ba\\
\;\ba^\d\end{array}\right)=S^\d\left(\begin{array}{cc} \ba\\
\;\ba^\d\end{array}\right),
\en{EW7}
due to the identities of Eq.  (\ref{EW6}) and the derivatives from those, which
lead to the matrix identity
\be
\left(\begin{array}{cc}U & V^*\\ V & U^*\end{array}
\right)\left(\begin{array}{cc}U^\d & V^\d\\ \wtilde{V} & \wtilde{U}\end{array}
\right) = \left(\begin{array}{cc}I& 0\\ 0 & I\end{array} \right).
\en{EW8}
By interchanging the order in the product of the matrices in Eq. (\ref{EW8}), we
obtain also the following relations
\be
U^\d U + V^\d V = I, \quad \wtilde{U} V + \wtilde{V} U = 0,
\en{EW9}
which are  equivalent to the relations (\ref{EW6}).

 The unitary matrix $S$ which gives a
canonical transformation must also satisfy the following property $\tau S \tau  =
S^*$, see  Eq. (\ref{EW5}). Hence there is such matrix $\Omega$ that
\be
S = \left(\begin{array}{cc}U & V^*\\ V & U^*\end{array} \right) = e^{i\Omega},\quad
\Omega^\d = \Omega,\quad \tau\Omega\tau = -\Omega^*.
\en{EW15}
This is the most general form of a linear canonical transformation of the fermion
creation and annihilation  operators.

Let us now consider what constraints on the transformation (\ref{EW5}) imposes
diagonalization by this of the Hamiltonian (\ref{EW3}). First of all, we must
obtain
\be
H = \sum_\mu \lambda_\mu b^\d_\mu b_\mu+\frac{\mathrm{Tr}(A)}{2}\equiv
\wtilde{\bb}^\d \Lambda\bb+\frac{\mathrm{Tr}(A)}{2},\quad \Lambda =
\mathrm{diag}(\lambda_1,\ldots,\lambda_N).
\en{EW10}
Then, calculating the commutators with the fermionic operators, we have
\[
[a_\mu, H] = \sum_\nu(A_{\mu\nu}a_\nu -B^*_{\mu\nu}a^\d_\nu),\quad [a^\d_\mu, H] =
\sum_\nu(-A^*_{\mu\nu}a^\d_\nu +B_{\mu\nu}a_\nu),
\]
i.e., in the matrix form
\be
[\ba,H] = A\ba-B^*\ba^\d,\quad [\ba^\d,H] = (-A^*\ba^\d + B\ba).
\en{EW11}
On the other hand, using the transformation and the form of the Hamiltonian
(\ref{EW10}), we have
\[
[\ba,H] = [U\bb+V^*\bb^\d,H] = U\Lambda\bb-V^*\Lambda\bb^\d,
\]
\be
[\ba^\d,H] = [U^*\bb^\d+V\bb,H] = V\Lambda\bb-U^*\Lambda\bb^\d.
\en{EW12}
Expressing the $a$-operators through $b$-ones in the r.h.s. of Eq. (\ref{EW11}),
using the r.h.s. of Eq. (\ref{EW12}) and  the linear independence of the creation
and annihilation operators\footnote{One can see this fact by employing the inner
products with the Fock space vectors $\vr{0}$ and $\vr{1}$ for each mode
$j=1,\ldots,N$, i.e. $\langle0|a_\mu\vr{1}=1$ while $\langle0|a^\d_\mu\vr{1}=0$.}
we arrive at the eigenvalue problem for the Hessian matrix (\ref{EW4}):
\be
\left(\begin{array}{cc}A &B^\d\\ B & -A^*\end{array}
\right)\left(\begin{array}{cc}U & V^*\\ V & U^*\end{array}\right) =
\left(\begin{array}{cc}U & V^*\\ V & U^*\end{array}\right)\left(\begin{array}{cc}\Lambda &0\\
0 & -\Lambda\end{array} \right).
\en{EW13}
It is seen that the diagonalization problem always has  a solution, since the
Hessian is Hermitian (hence the real eigenvalues) while its properties guarantee
that the eigenvalues always come in pairs $(\lambda_\mu,-\lambda_\mu)$ and  the
relation between the eigenvectors implied in Eq. (\ref{EW13}) holds. Indeed,
consider the eigenvector $\vr{\lambda_\mu}$\footnote{For notational convenience, we
will use the Dirac notations also for the eigenvectors of matrices: the
vector-columns will be the ket-vectors and the vector-rows -- the bra-vectors.}
corresponding to the eigenvalue $\lambda_\mu\ne0$. Due to properties (\ref{EW4})
and that $\tau^2=I$ we have:
\be
\mathcal{H}\vr{\lambda_\mu} = \lambda_\mu\vr{\lambda_\mu},\quad
\mathcal{H}\left(\tau\vr{\lambda_\mu}^*\right) =
-\lambda_\mu\left(\tau\vr{\lambda_\mu}^*\right),
\en{EW14}
which is precisely the relation expressed in Eq. (\ref{EW13}) between the
eigenvectors, i.e.
\[
\vr{\lambda_\mu} = \left(\begin{array}{c} U_{\cdot\mu}\\
V_{\cdot\mu}\end{array} \right),\quad \vr{-\lambda_\mu} = \left(\begin{array}{c} V^*_{\cdot\mu}\\
U^*_{\cdot\mu}\end{array}\right).
\]
The above consideration could fail when the eigenvalue is equal to zero. However,
since the non-zero eigenvalues come in pairs and the Hessian matrix is traceless
(see Eq. (\ref{EW3})), an even number of linearly independent eigenvectors
corresponds to zero eigenvalue. In the case when two (or any other, necessarily
\textit{even} number) of them satisfies the property $\tau\vr{X}^* = \vr{X}$ and
$\tau\vr{Y}^* = \vr{Y}$ one can take the vectors $\vr{Z_\pm} =  \vr{X}\pm i \vr{Y}$
which satisfy the required  property $\tau\vr{Z_+}^* = \vr{Z_-}$, needed for the
unitary transformation $S$, diagonalizing the Hessian,  to be also a canonical one.

%%%%%%%%%%%%%%%% EVOLUTION FERMIONS %%%%%%%%%%%%%%%%%%%%%%%%%%%%%%%%%%%%%%%%%

Consider now the quantum evolution described by a fermionic quadratic Hamiltonian
\be
\hat{H} = \frac12\left(\begin{array}{cc}\wtilde{\ba}^\d,\wtilde{\ba}\end{array}
\right) \left(\begin{array}{cc}E & F^\d\\ F & -E^*\end{array}
\right)\left(\begin{array}{c} \ba\\ \;\ba^\d\end{array} \right),
\en{EW16}
where $E^\d = E$ and $\wtilde{F} = -F$ as required. In the Heisenberg picture,  the
evolution  by a unitary operator $\mathcal{U}(t) = e^{-it\hat{H}/\hbar}$ is
transferred to  the creation and annihilation operators, i.e.
\be
\ba(t) = e^{it\hat{H}/\hbar}\ba e^{-it\hat{H}/\hbar}, \quad \ba^\d(t) =
e^{it\hat{H}/\hbar}\ba^\d e^{-it\hat{H}/\hbar},
\en{EW17}
which is obviously also a canonical transformation. Moreover, it is also a linear
canonical transformation and  can be put in the form of Eq. (\ref{EW5}). To arrive
at the required transformation matrix, let us  use by the well-known identity
\be
e^BAe^{-B} = \sum_p\frac{[B,A]^{(p)}}{p!},
\en{ID}
where $[B,A]^{(p)} \equiv [B,[B,\ldots,[B,A]\ldots],],]$ is the successive
commutator  of $B$ with $A$ of order $p$. We must  compute the successive
commutators of the Hamiltonian $\hat{H}$ with the creation and annihilation
operators $a_\mu, a^\d_\mu$. By using the property of commutator of fermionic
operator with an even number of operators, Eq. (\ref{EQ69}) of section \ref{secI6},
we obtain
\be
[\hat{H},a_\mu] = \sum_{\nu}(-E_{\mu\nu}a_\nu+F^*_{\mu\nu}a^\d_\nu),\quad
[\hat{H},a^\d_\mu] = \sum_{\nu}(E^*_{\mu\nu}a^\d_\nu-F_{\mu\nu}a_\nu).
\en{comm_H}
Introducing the vector-operator  $\mathcal{A} =(a_1,\ldots,a_N,a^\d_1,\ldots,a^\d_N) $   
we have from Eq. (\ref{comm_H}) 
\[
[\hat{H},\mathcal{A}_n] = -
\sum_{\ell}\hat{\mathcal{H}}_{n\ell}\mathcal{A}_\ell,\quad
[\hat{H},[\hat{H},\mathcal{A}_n]] =
\sum_{\ell}{\mathcal{H}}_{n\ell}\sum_{m}{\mathcal{H}}_{\ell
m}\mathcal{A}_m, \ldots,
\]
 where \be
{\mathcal{H}} \equiv \left(\begin{array}{cc}E & F^\d\\ F & -E^*\end{array}
\right).
\en{EW18}
In the c onsise from the above commutators read
\[
[\hat{H},\mathcal{A}_n]^{(p)} =
\sum_\ell\left\{(-{\mathcal{H}})^p\right\}_{n\ell}\mathcal{A}_\ell
\]
which implies that
\be
\left(\begin{array}{c} \ba(t)\\ \;\ba^\d(t)\end{array} \right)=
e^{it\hat{H}/\hbar}\left(\begin{array}{c} \ba\\ \;\ba^\d\end{array}
\right)e^{-it\hat{H}/\hbar} = e^{-it{\mathcal{H}}/\hbar} \left(\begin{array}{c} \ba\\
\;\ba^\d\end{array} \right).
\en{EW19}
The r.h.s. of this equation  gives a  linear canonical transformation by the
properties of the Hessian ${\mathcal{H}}$ (cf. with Eq. (\ref{EW15})). By
comparing Eqs. (\ref{EW19}) and (\ref{EW15}) we see that a canonical transformation
linear in the creation and annihilation operators can be expressed as a result of a
temporal evolution given by another quadratic Hamiltonian with the Hessian provided
by the exponential representation of the transformation  as in Eq. (\ref{EW15}):
${\mathcal{H}} = \hbar\Omega$ (for $t=1$).
%%%%%%%%%%%%%%%%%%%%%%%%%%%%%%%%%%%%%%%%%%%%%%%%%%%%%%%%%%%%%%%%%%%%%%%%%%%%%%%%%%%%%%%%%%%%%%%%%%%

\section{Diagionalization of the Hamiltonian quadratic in the boson operators}
\label{secII2}

Let us start with considering the quantum evolution governed by a Hamiltonian
quadratic in bosonic operators,
\be
\hat{H} = \frac12\left(\begin{array}{cc}\wtilde{\ba}^\d,\wtilde{\ba}\end{array}
\right) \left(\begin{array}{cc}E & F^*\\ F & E^*\end{array}
\right)\left(\begin{array}{c} \ba\\ \;\ba^\d\end{array} \right).
\en{EW20}
Here we have used the bosonic commutation relations to see that a general quadratic
Hamiltonian,
\[
\hat{H} = \sum_{\mu\nu}E_{\mu\nu}a^\d_\mu a_\nu +
\frac12\sum_{\mu\nu}F_{\mu\nu}a_\mu a_\nu+\frac12\sum_{\mu\nu}F^*_{\mu\nu}a^\d_\mu
a^\d_\nu,
\]
implies that  $E^\d=E$ and $\wtilde{F}= F$ and,  by using
\[
\wtilde{\ba}^\d E\ba =  \wtilde{\ba}\wtilde{E}\ba^\d - \mathrm{Tr}(E)=
\wtilde{\ba}E^*\ba^\d - \mathrm{Tr}(E),
\]
after dropping a scalar $-\mathrm{Tr}(E)/2$, can be cast in the form of Eq.
(\ref{EW20}). First, note that the Hessian matrix associated with the Hamiltonian
(\ref{EW20})  satisfies the usual properties of a Hessian of a classical
Hamiltonian system:
\be
 {\mathcal{H}} \equiv \left(\begin{array}{cc}E & F^*\\ F & E^*\end{array}
\right),\quad {\mathcal{H}}^\d = {\mathcal{H}},\quad
\tau{\mathcal{H}}\tau = {\mathcal{H}}^*.
\en{EW21}
This analogy is essential for the below discussed diagonalization of quadratic
bosonic Hamiltonian.

Consider the  unitary evolution operator $\mathcal{U}(t) = e^{-it\hat{H}/\hbar}$
which, in the Heisenberg picture, induces a canonical transformation
\be
\ba(t) = e^{it\hat{H}/\hbar}\ba \,e^{-it\hat{H}/\hbar}, \quad \ba^\d(t) =
e^{it\hat{H}/\hbar}\ba^\d e^{-it\hat{H}/\hbar}.
\en{EW22}
Similarly as in the fermion case, introducing the vector-operator  $\mathcal{A}
=(a_1,\ldots,a_N,a^\d_1,\ldots,a^\d_N) $   we  obtain 
the successive  commutators $[\hat{H},\mathcal{A}_n]$ and  arrive at the relation
\[
[\hat{H},\mathcal{A}_n]^{(p)} =
\sum_\ell\left\{(-J{\mathcal{H}})^p\right\}_{n\ell}\mathcal{A}_\ell,\quad
J\equiv \left(\begin{array}{cc}I_{N\times N} & 0\\ 0 & -I_{N\times N} \end{array}
\right),
\]
i.e. $J$ is the block diagonal matrix, where each block has the dimension $N\times
N$. Thus, using Eq. (\ref{ID}) we obtain the identity
\be
\left(\begin{array}{c} \ba(t)\\ \;\ba^\d(t)\end{array} \right)=
e^{it\hat{H}/\hbar}\left(\begin{array}{c} \ba\\ \;\ba^\d\end{array}
\right)e^{-it\hat{H}/\hbar} = e^{-itJ{\mathcal{H}}/\hbar} \left(\begin{array}{c} \ba\\
\;\ba^\d\end{array} \right).
\en{EW23}
We will see below that, similar to the fermion case,  the matrix on the r.h.s. of
this equation, with the Hessian of the form of Eq.  (\ref{EW21}), gives a canonical
transformation linear in bosonic operators. Note that in the bosonic case, as
distinct from the fermionic one, the unitary operator evolution is expressed by the
transformation with  a non-unitary matrix (hence  the possibility of instabilities
in the Hamiltonian dynamics describing the bosonic quasiparticles).

%%%%%%%%%%%%%%%%%% GENERAL CANONICAL TRANSFORAMTION %%%%%%%%%%%%%%%%%%%%%%%%%%%%%%%%%%%%%%%%%%%%%%%%

Let us now return to the diagonalization problem. First of all, consider a general
linear transformation of the bosonic operators
\be
\left(\begin{array}{cc} \ba\\ \;\ba^\d\end{array} \right)= \left(\begin{array}{cc}U
& V^*\\ V & U^*\end{array} \right) \left(\begin{array}{cc} \bb\\
\;\bb^\d\end{array}\right)\equiv S\left(\begin{array}{cc} \bb\\
\;\bb^\d\end{array}\right).
\en{EW24}
The commutation relations then give the following constraints
\[
[a_\mu,a_\nu] =
\sum_{\alpha,\beta}[U_{\mu\alpha}b_\alpha+V^*_{\mu\alpha}b^\d_\alpha,U_{\nu\beta}b_\beta+V^*_{\nu\beta}b^\d_\beta]
=\sum_{\alpha}(U_{\mu\alpha}V^*_{\nu\alpha}-V^*_{\mu\alpha}U_{\nu\alpha})=0,
\]
\[
[a_\mu,a^\d_\nu] =
\sum_{\alpha,\beta}[U_{\mu\alpha}b_\alpha+V^*_{\mu\alpha}b^\d_\alpha,U^*_{\nu\beta}b^\d_\beta+V_{\nu\beta}b_\beta]
=\sum_{\alpha}(U_{\mu\alpha}U^*_{\nu\alpha}-V^*_{\mu\alpha}V_{\nu\alpha})=\delta_{\mu,\nu},
\]
i.e., in the matrix form, we have obtained the relations
\be
UU^\d-V^*\wtilde{V}=I,\quad UV^\d-V^*\wtilde{U}=0.
\en{EW25}
These conditions also guarantee   the existence of the inverse transformation to
that of Eq. (\ref{EW24}), in this case we have (cf. with Eq. (\ref{EW7}))
\be
\left(\begin{array}{cc} \bb\\ \;\bb^\d\end{array} \right)=
\left(\begin{array}{cc}U^\d &
-V^\d\\ -\wtilde{V} & \wtilde{U}\end{array} \right) \left(\begin{array}{cc} \ba\\
\;\ba^\d\end{array}\right)=JS^\d J\left(\begin{array}{cc} \ba\\
\;\ba^\d\end{array}\right).
\en{EW26}
Eq. (\ref{EW26}) can be verified by using the identities  (\ref{EW25}) and the ones
derived from those, which  lead to
\be
\left(\begin{array}{cc}U & V^*\\ V & U^*\end{array}
\right)\left(\begin{array}{cc}U^\d & -V^\d\\ -\wtilde{V} & \wtilde{U}\end{array}
\right) = \left(\begin{array}{cc}I& 0\\ 0 & I\end{array} \right),
\en{EW27}
i.e. $S^{-1} = JS^\d J$.  By interchanging the order of the matrices in the product
in Eq. (\ref{EW26}) we obtain also
\be
U^\d U - V^\d V = I, \quad \wtilde{U} V - \wtilde{V} U = 0.
\en{EW28}
Note that the relations (\ref{EW28}) are equivalent to those of Eq. (\ref{EW25}),
i.e.  one set of relations leads to the other.

Note   the following property of the bosonic canonical transformation (\ref{EW24}),
the matrix $U$ is invertible, i.e. $\mathrm{det}(U)\ne0$. Indeed, the first
relation in Eq. (\ref{EW28}) excludes the possibility of non-trivial vectors in the
kernel of  matrix $U$, since $U|x\rangle=0$ leads to a contradiction:
\mbox{$0>-\langle x|V^\d V|x\rangle = \vl{x}x\rangle >0$.} This is in contrast to
the fermion case, where both $U$ and $V$ can be zero (the trivial exchange of the
creation and annihilation operators is an example of a canonical transformation in
the fermionic case).

Let us show that any canonical transformation (\ref{EW24}) can be represented in
the same form as on the r.h.s. of Eq. (\ref{EW23}), thus proving that any linear
canonical transformation can be represented by a unitary evolution with some
quadratic Hamiltonian, exactly as in the fermion case. To this end we have to show
that the transformation matrix $S$ of Eq. (\ref{EW24}) can be cast in the
exponential form, that is
\be
S = \left(\begin{array}{cc}U & V^*\\ V & U^*\end{array} \right) =
e^{-iJ{\mathcal{H}}},
\en{EXPB}
with some  matrix ${\mathcal{H}}$, i.e. having the properties of a Hessian:
${\mathcal{H}}^\d = {\mathcal{H}}$ and $\tau{\mathcal{H}}\tau =
{\mathcal{H}}^*$. But the latter properties, together with the identity
$J\tau=-\tau J$, are precisely the properties which guarantee that $\tau S\tau =
S^*$, i.e. the right form of matrix $S$ as on the l.h.s. of Eq. (\ref{EXPB}), and
$S^{-1} = JS^\d J$, Eq. (\ref{EW27}), which is equivalent to the transformation
being canonical, i.e. the conditions (\ref{EW25}) being satisfied. Moreover, one
can easily verify by direct calculation that the canonical transformations
(\ref{EW24}) form a group, i.e. a composition of two canonical transformations is
also a canonical transformation.

%%%%%%%%%%%%%%%%%%%%%% DIAGONALIZATION PART %%%%%%%%%%%%%%%%%%%

Consider now the problem of diagonalization of a quadratic bosonic Hamiltonian
\[
H = \sum_{\mu\nu}A_{\mu\nu}a^\d_\mu a_\nu + \frac12\sum_{\mu\nu}B_{\mu\nu}a_\mu
a_\nu+\frac12\sum_{\mu\nu}B^*_{\mu\nu}a^\d_\mu a^\d_\nu
\]
\be
= \frac12\left(\begin{array}{cc}\wtilde{\ba}^\d,\wtilde{\ba}\end{array} \right)
\left(\begin{array}{cc}A & B^*\\ B & A^*\end{array} \right)\left(\begin{array}{c}
\ba\\ \;\ba^\d\end{array} \right) - \frac12\mathrm{Tr}(A),
\en{EW29}
where $A^\d = A$ and $\wtilde{B}=B$, as required. By the same procedure as in the
fermionic case, i.e. by using the commutation relations and the diagonalized form
of the Hamiltonian (\ref{EW29}),
\be
H = \sum_\mu \lambda_\mu b^\d_\mu b_\mu+K\equiv \wtilde{\bb}^\d \Lambda\bb+K,\quad
\Lambda = \mathrm{diag}(\lambda_1,\ldots,\lambda_N),
\en{EW30}
where $K$ is a scalar constant due to the non-commutation of the creation and
annihilation operators (it will be computed below), we obtain the following
equations
\be
AU+B^*V = U \Lambda, \quad BU +A^*V = -V\Lambda.
\en{EW31}
These lead to a  \textit{non-Hermitian} eigenvalue problem with the eigenvectors
being the columns of the   canonical transformation diagonalizing the Hamiltonian
(\ref{EW29}):
\be
J\mathcal{H} \left(\begin{array}{cc}U & V^*\\ V & U^*\end{array} \right)=
\left(\begin{array}{cc}U & V^*\\ V & U^*\end{array}
\right)\left(\begin{array}{cc}\Lambda& 0\\ 0 & -\Lambda\end{array} \right),
\en{EW32}
where $J$ is defined above Eq. (\ref{EW23}), while $\mathcal{H}$ is the  Hessian of
the Hamiltonian, i.e.
\be
\mathcal{H} = \left(\begin{array}{cc}A & B^*\\ B & A^*\end{array} \right).
\en{EW33}

First of all, the eigenvalue problem for the matrix $J\mathcal{H}$, where
$\mathcal{H}$ is Hermitian and  satisfies the property $\tau\mathcal{H}\tau =
\mathcal{H}^*$ is related to the linear stability problem of a classical
Hamiltonian system whose Hessian at the solution to be studied for stability is
exactly $\mathcal{H}$ (see below). Let us discuss some properties of the eigenvalue
problem (\ref{EW33}).  The eigenvalues come in quartets
$\{\lambda_\mu,-\lambda_\mu,\lambda^*_\mu,-\lambda^*_\mu\}$. This is easily seen
from the properties of $\mathcal{H}$ and that $J\tau = -\tau J$:
\be
J\mathcal{H}\vr{X} = \lambda\vr{X},\quad J\mathcal{H}\left(\tau\vr{X}^*\right) =
-\lambda^*\left(\tau\vr{X}^*\right),
\en{EW34}
 while from the existence of the left-eigenvector
$\quad \vl{Y}J\mathcal{H} = \lambda\vl{Y}$ we get
\[
\quad J\mathcal{H}\left(J\vr{Y}\right) = \lambda^*\left(J\vr{Y}\right),\quad
J\mathcal{H}\left(\tau J \vr{Y}^*\right) = -\lambda\left(\tau J\vr{Y}^*\right).
\]
The inner product which corresponds to the  normalization condition of the
eigenvectors, required by the canonical transformation, i.e. $U^\d U -V^\d V = I$
where we use $\vr{X_\mu} = (U_{\cdot\mu},V_{\cdot\mu})^T$, must be  defined as
$\langle X_\mu|J|X_\nu\rangle$ and  is indefinite. The other condition of Eq.
(\ref{EW28}) is the orthogonality condition in this inner product between the
eigenvectors corresponding to different eigenvalues, i.e.  the eigenvectors
$|X_\mu\rangle$ corresponding to $\lambda_\mu$ and $\tau|X_\nu\rangle^*$
corresponding to $-\lambda^*_\nu$. The eigenvectors corresponding to the eigenvalues
$\lambda_1$ and $\lambda_2$  are orthogonal if $\lambda_2\ne\lambda_1^*$, since
\[
\langle X_1|\mathcal{H}| X_2\rangle = \lambda_2\langle X_1|J| X_2\rangle =
\lambda_1^*\langle X_1|J|X_2\rangle.
\]
For real eigenvalues this reduces to the orthogonality of the eigenvectors
corresponding to different eigenvalues. In this case we have also
\[
\langle X_\pm|\mathcal{H}|X_\pm\rangle =\pm\lambda\langle X_\pm|J|X_\pm\rangle,
\]
hence one of the eigenvectors,  i.e. corresponding to either $\lambda_\mu$ or
$-\lambda_\mu$, has   positive  Krein index $\kappa$ defined as
\be
\kappa(|X\rangle)= \mathrm{sign}(\langle X|J|X \rangle)
\en{Krein}
since the l.h.s. is positive) and can be used as a column in the matrix $(U,V)^T$.
Thus, under the condition that the eigenvalues are semisimple, i.e. the number of
the corresponding eigenvectors is equal to the algebraic multiplicity of the
eigenvalue\footnote{The matrix $J\mathcal{H}$ is not normal, unless it is
block-diagonal, hence its normal Jordan form is  not diagonal, in general, and
there may be generalized eigenvectors -- such is the case of the  zero eigenvalue,
see below.}, one can always find the first $N$ columns of the diagonalizing
canonical transformation in Eq. (\ref{EW32}). Then, the relation (\ref{EW34}) tells
us that the matrix $S$, diagonalizing $J\mathcal{H}$ is also a canonical
transformation, as in Eq. (\ref{EW32}). One special case, when the eigenvalue is
not semisimple, is the case of zero eigenvalue. It will be considered separately
below.

%%%%%%%%%%%%%%%%%%%%% Relation to Stability of Hamiltonian flow %%%%%%%%%%%%%%

Let us now verify that the canonical transformation of Eq. (\ref{EW32}), i.e.
satisfying the properties (\ref{EW31}), indeed diagonalizes the Hamiltonian
(\ref{EW29}) and compute the scalar constant $K$ which has appeared in the diagonal
form (\ref{EW30}). Using Eqs. (\ref{EW29}), (\ref{EW24}) and (\ref{EW32}) we obtain
\[
H = \frac12\left(\begin{array}{cc}\wtilde{\bb}^\d,\wtilde{\bb}\end{array}
\right)\left(\begin{array}{cc}U^\d & V^\d\\ \wtilde{V} & \wtilde{U}\end{array}
\right) \mathcal{H}\left(\begin{array}{cc}U & V^*\\ V & U^*\end{array}
\right)\left(\begin{array}{c} \bb\\ \;\bb^\d\end{array} \right) -
\frac12\mathrm{Tr}(A)
\]
\[
=\frac12\left(\begin{array}{cc}\wtilde{\bb}^\d,\wtilde{\bb}\end{array}
\right)J\left(\begin{array}{cc}\Lambda& 0\\ 0 & -\Lambda\end{array}
\right)\left(\begin{array}{c} \bb\\ \;\bb^\d\end{array} \right) -
\frac12\mathrm{Tr}(A)
\]
\[
=\wtilde{\bb}^\d \Lambda \bb + \frac12\mathrm{Tr}(\Lambda-A).
\]
Hence, using Eq. (\ref{EW32}) to obtain $A = U\Lambda U^\d + V^*\Lambda \wtilde{V}$
we have
\[
K = \frac12\mathrm{Tr}(\Lambda-A) = \frac12\mathrm{Tr}\{(I - U^\d U -
\wtilde{V}V^*)\Lambda\} =\frac12\mathrm{Tr}\{(I - U^\d U - V^\d V)\Lambda\}
\]
\be
=- \mathrm{Tr}(V^\d V\Lambda) = -\sum_\mu\lambda_\mu\sum_\nu|V_{\nu\mu}|^2.
\en{EW36}

The diagonalization of $J\mathcal{H}$ is intimately related to the properties of
the quadratic form $\frac12\langle X|\mathcal{H}|X\rangle$ which is a Lyapunov
(energy) function for the classical Hamiltonian system $i\dot{X} = J\mathcal{H} X$,
written in the complex canonical coordinates: $X =
(z_1,\ldots,z_N,z_1^*,\ldots,z_N^*)$. The Hamiltonian flow is decoupled into
invariant (independent) flows corresponding to the subspaces of the quartets of
eigenvalues. It can be shown that the quadratic form $\langle
X|\mathcal{H}|X\rangle$ has even numbers of positive and negative squares over each
subspace corresponding to real eigenvalues $\mathrm{Im}(\lambda_\mu)=0$ (but not
necessarily being definite in any of these), while it has equal number of positive
and negative squares over each  subspace corresponding to the eigenvalues having
nonzero imaginary part, $\mathrm{Im}(\lambda_\mu)\ne0$.

Therefore,   a  sufficient condition  for the matrix $J\mathcal{H}$ to have real
eigenvalues, i.e. the corresponding Hamiltonian flow to be stable,  is the
quadratic form $\langle X|\mathcal{H}|X\rangle$   be definite. This is a very crude
criterion, since the quadratic form can be indefinite over the whole space, but
definite in each of the invariant subspaces which property is sufficient  for  the
matrix $J\mathcal{H}$ to have  real eigenvalues  only. We still have  to show that
the definiteness of $\mathcal{H}$ excludes the   generalized eigenvectors of
$J\mathcal{H}$ (we have used that the eigenvalues are semisimple). Consider   a
definite $\mathcal{H}$, then the matrix $B = \mathrm{sign}(\mathcal{H})
J\mathcal{H} J $ is positive. Suppose that there exists a generalized eigenvector
$\vr{X_1}$ for the eigenvalue $\lambda$ and the eigenvector $\vr{X_0}$, i.e.
\[
J\mathcal{H}  |X_1\rangle = \lambda\vr{X_1} +\vr{X_0}, \quad J\mathcal{H}
|X_0\rangle = \lambda\vr{X_0}. \] We have then
\[
J\mathcal{H} J\mathcal{H}\vr{X_1} = \lambda^2\vr{X_1} + 2\lambda\vr{X_0},
\]
hence
\[
B^\frac12 \mathcal{H} B^\frac12 |\wtilde{X}_1\rangle = \lambda^2\vr{\wtilde{X}_1}
+\vr{\wtilde{X}_0}, \quad  B^\frac12 \mathcal{H} B^\frac12 |\wtilde{X}_0\rangle =
\lambda^2\vr{\wtilde{X}_0},
\]
where
\[
\vr{\wtilde{X}_1} \equiv B^{-\frac12}\vr{\wtilde{X}_1}, \quad \vr{\wtilde{X}_0} \equiv 2\lambda B^{-\frac12}\vr{\wtilde{X}_0},
\]
which is a contradiction, since the  matrix $B^\frac12 \mathcal{H} B^\frac12 $ is
Hermitian and cannot have generalized eigenvectors.

%%%%%%%%%%%%%%%%%%%%%%%%%%%%%%%%%%%%%%%%%%%%%%%%%%%%%%%%%%%%%%%%%%%%%%%%%%%%%%%%%%%%%%%%%%

\subsection{Diagonalization of the quadratic bosonic Hamiltonian possessing
a zero mode}
\label{subII2A}

Let us consider now the special case, when there  is a zero eigenvalue of the
matrix $J\mathcal{H}$ (\ref{EW32}) used in the diagonalization of the quadratic
bosonic Hamiltonian (\ref{EW29}).  Let us assume that there is just one eigenvector
$|X_0\rangle$,  also called the zero mode,   corresponding to the zero eigenvalue,
i.e. $J\mathcal{H}|X_0\rangle=0$. Then $\tau|X_0\rangle^* =
e^{i\theta}|X_0\rangle$, with some phase $\theta$ (evident from
$\tau(\tau|X_0\rangle^*)^* = |X_0\rangle$).  Note that the phase $\theta$ can be
set to zero by multiplication of the zero mode by $e^{i\theta/2}$. We have then
\be
|X_0\rangle = \left(\begin{array}{c}G\\ G^*\end{array} \right), \quad
J\mathcal{H}|X_0\rangle=0,
\en{EW38}
where $G$ is a column. This fact disallows one to use the zero mode as one of the
columns in the construction of the transformation matrix (\ref{EW32}), otherwise it
will not have  the inverse and hence, the boson commutation relations will not be
satisfied by the new operators.

However the remedy is in the  eigenvalue problem  itself, which is not Hermitian
and allows the existence of the generalized eigenvectors. There is at least one of
them. Indeed, we have for the first generalized eigenvector the following problem
\be
J\mathcal{H}|X_1\rangle = |X_0\rangle.
\en{EW39}
Due to the fact that $\langle X_0|\mathcal{H} = 0$  and $\langle X_0|J|X_0\rangle =
0$, such a  generalized eigenvector $|X_1\rangle$ exists  by the Fredholm
alternative. Note that Eqs. (\ref{EW38}) and (\ref{EW39}) give $\tau|X_1\rangle^* =
- |X_1\rangle$. i.e. the general form of the generalized eigenvector reads
\be
|X_1\rangle = \left(\begin{array}{c}F\\ -F^*\end{array} \right),
\en{EW40}
where $F$ is a column.

We assume the simplest case, when  there are   no higher generalized eigenvectors,
by assuming the  condition $\langle X_0|J|X_1\rangle  =\langle
X_1|\mathcal{H}|X_1\rangle \ne0$ (the product is  then positive), which disallows
the existence  of the higher eigenvector $|X_2\rangle$, i.e. the corresponding
equation
\[
J\mathcal{H}|X_2\rangle = |X_1\rangle
\]
by the Fredholm alternative has no solutions.  Below the vectors $|X_0\rangle$ and
$|X_1\rangle$ are assumed to be normalized such that $\langle X_0|J|X_1\rangle=1$
(since their product is always positive, the normalization does not change the form
of the eigenvectors given by  Eqs. (\ref{EW38}) and (\ref{EW40})).  Introduce the
following normalized vector
\be
|Z \rangle = \left(\begin{array}{c}Z_U\\ Z_V\end{array} \right) =
\alpha|X_0\rangle+|X_1\rangle, \quad \alpha +\alpha^*=1,
\en{EW41}
i.e. that $\langle Z|J|Z\rangle = Z^\d_UZ_U -Z^\d_V Z_V = 1$. Now we can construct
the transformation matrix $S$  of Eq. (\ref{EW24}) having the required properties:
\be
S = \left(\begin{array}{cc}Z_U , U_\perp & Z^*_V , V^*_\perp\\ Z_V, V_\perp &
Z^*_U, U^*_\perp\end{array} \right)\equiv \left(\begin{array}{cc}  U  & V^* \\ V &
U^* \end{array} \right),
\en{EW42}
where the matrices $U_\perp$ and $V_\perp$ correspond to the non-zero eigenvalues
of $J\mathcal{H}$. Due to the orthogonality of the eigenvectors  corresponding to
distinct eigenvalues and the normalization of $|Z\rangle$ we have $S^{-1} = JS^\d
J$ as it should be. On the other hand, by construction we have  $\tau S\tau = S^*$.
The eigenvector decomposition of the transformation matrices $S$ and $S^{-1}$
reads:
\be
S = \left(|Z\rangle, \ldots, |X_\mu\rangle, \ldots, \tau|Z\rangle^*, \ldots, \tau|X_\mu\rangle^*, \ldots\right), \quad
S^{-1}= \left(\begin{array}{c}\langle Z|J\\ \vdots\\ \langle X_\mu|J\\
 \vdots \\ -\langle Z|^*\tau J\\ \vdots \\ -\langle X_\mu|^*\tau J \\ \vdots\end{array} \right),
\en{Svec}
where we have used the property $S^{-1} = JS^\d J$ and the  diagonal structure of
$J$, which lead to the minus sign at $\langle X_\mu|^*\tau J$.  Observe that the
minus sign in the inverse transformation matrix is at the vectors having negative
Krein index (\ref{Krein}), due to our selection as the vectors $|X_\mu\rangle$
those which have positive Krein index (since we had to satisfy the condition  $U^\d
U = V^\d V = I$). From Eq. (\ref{Svec}) and  the general theory of Jordan normal
form of a matrix (using that $\langle X_0|J|X_1\rangle = 1$ for vectors of the
zero eigenvalue subspace), one easily derives the representation (note the
appearance of the  Krein index in the formula below)
\[
J\mathcal{H}  = \sum_{\lambda}\lambda  \frac{|X_\lambda\rangle\langle
X_\lambda|J}{\langle X_\lambda|J|X_\lambda\rangle} + \frac{|X_0\rangle\langle
X_0|J}{\langle X_0|J|X_1\rangle} = \sum_{\lambda}\lambda
\kappa(|X_\lambda\rangle)|X_\lambda\rangle\langle X_\lambda|J + |X_0\rangle\langle
X_0|J
\]
\be
= S\left(\begin{array}{cc}\Lambda& 0\\
0 & -\Lambda\end{array} \right)S^{-1} +  |X_0\rangle\langle X_0|J,
\en{EW43}
where $\Lambda = \mathrm{diag}(0,\lambda_2,\ldots,\lambda_{N})$, $\lambda_\mu\ne0$
for $\mu\ne1$, $|X_\lambda\rangle $ is the  eigenvector  corresponding to $\lambda$
and the addition on the r.h.s. is the nilpotent matrix $N_0$, such that
$N_0|X_0\rangle = 0$ and $N_0|X_1\rangle = |X_0\rangle$, i.e.  giving the Jordan
block corresponding to the zero eigenvalue (of multiplicity 2 in our case). From
Eq. (\ref{EW43}) the structure of the Hessian matrix associated with the
Hamiltonian (\ref{EW29}) appearing  under the canonical transformation (\ref{EW42})
can be calculated. First of all, using the orthogonality  properties of the
eigenvectors corresponding to different eigenvalues and that $\langle
Z|J|X_0\rangle = 1$ and $\langle Z|^*\tau J|X_0\rangle = -1$ one can obtain the
identity
\be
\langle X_0|JS \left(\begin{array}{c} \bb\\ \;\bb^\d\end{array}  \right) =
(1,0,\ldots,0,-1,0,\ldots,0)\left(\begin{array}{c} \bb\\ \;\bb^\d\end{array}
\right) = b_1 - b^\d_1.
\en{X0b}
Using this identity and Eqs. (\ref{EW38})--(\ref{EW43}) we get
\[
H = \frac12\left(\begin{array}{cc}\wtilde{\bb}^\d,\wtilde{\bb}\end{array}
\right)S^\d \mathcal{H} S\left(\begin{array}{c} \bb\\ \;\bb^\d\end{array} \right) -
\frac12\mathrm{Tr}(A)
\]
\[
=\frac12\left(\begin{array}{cc}\wtilde{\bb}^\d,\wtilde{\bb}\end{array}
\right)\left\{\left(\begin{array}{cc}\Lambda& 0\\ 0 & \Lambda\end{array}
\right)+S^\d J|X_0\rangle\langle X_0|JS \right\}\left(\begin{array}{c} \bb\\
\;\bb^\d\end{array} \right) - \frac12\mathrm{Tr}(A)
\]
\be
= \wtilde{\bb}^\d \Lambda \bb +\frac12(b^\d_1 - b_1)(b_1 - b^\d_1)
+\frac{\mathrm{Tr}(\Lambda-A)}{2}.
\en{EW44}
Note the appearance of the non-diagonalized part expressed through the new operators
$b_1$ and $b^\d_1$. It is easy to see that this addition represents the simplest
one-mode bosonic Hamiltonian   quadratic in the boson operators,
\[
H_1\equiv \frac12(b^\d_1 - b_1)(b_1 - b^\d_1) = \frac12\left(b^\d_1b_1 +b_1b^\d -b_1^2 -(b^\d_1)^2 \right),
\]
which cannot be diagonalized. Indeed, let us perform  the canonical change of
variables to the real observables $Q$ and $P$, defined as follows
\be
Q_1 \equiv \frac{b_1 +b^\d_1}{\sqrt{2}},\quad P_1\equiv \frac{b_1
-b^\d_1}{i\sqrt{2}},\quad [P_1,Q_1] = -i.
\en{EW45}
Then $H_1 = P^2_1$ which has a \textit{continuous spectrum} as distinct from a
one-mode quadratic Hamiltonian which is diagonalizable to the standard form $\sim
c^\d c$. This continuous spectrum is a consequence of the symmetry breaking in the
associated classical Hamiltonian system with the same  Hessian $\mathcal{H}$, a
feature which leads to the existence of the zero modes. Therefore, in the quantum
case, the classical symmetry breaking   leads to the appearance of  degrees of
freedom with a continuous spectrum.  Now,  we can write the Hamiltonian
(\ref{EW44}) in its final form
\be
H = P_1^2+ \sum_{\lambda_\mu\ne0}\lambda_\mu b^\d_\mu b_\mu
+\frac12\left[\sum_\mu\lambda_\mu-\mathrm{Tr}(A)\right].
\en{EW46}
Finally, let us express the canonical observables $P_1$ and $Q_1$ through the old
bosonic operators $a_\mu$ and $a^\d_\mu$.  From Eq. (\ref{X0b}) and the canonical transformation (\ref{EW24}) we get
\be
P_1 = -\frac{i}{\sqrt{2}}\langle X_0|J \left(\begin{array}{c}  \ba\\
\;\ba^\d\end{array} \right) =\frac{i}{\sqrt{2}}\sum_\mu\left(G_\mu a^\d_\mu -
G^*_\mu a_\mu \right).
\en{P1}
On the other hand, Eqs. (\ref{EW38}),
(\ref{EW40}),  (\ref{EW41}) and (\ref{Svec}) give
\[
b_1 =\langle Z|J\left(\begin{array}{c}\ba\\ \ba^\d\end{array}\right) = \sum_\mu
\left[(Z^*_U)_\mu a_\mu - (Z^*_V)_{\mu}a^\d_\mu\right]
\]
\be
 = \sum_\mu\left\{\left[\alpha^*G^*_\mu
+F^*_\mu\right]a_\mu+\left[F_\mu-\alpha^*G_\mu\right]a^\d_\mu\right\}.
\en{b1}
Therefore, by the definition (\ref{EW45}) and that $\alpha+\alpha^*=1$  we obtain
\be
Q_1 = \sqrt{2}\sum_\mu\left(F_\mu a^\d_\mu + F^*_\mu a_\mu \right) +
2\mathrm{Im}(\alpha)P_1= \sqrt{2}\langle X_1|J\left(\begin{array}{c}\ba\\
\ba^\d\end{array}\right) + 2\mathrm{Im}(\alpha)P_1.
\en{Q1}
The zero mode $|X_0\rangle$ and the generalized eigenvector $|X_1\rangle$
define the respective canonical operators.  Observe that there is some
arbitrariness in the definition of $Q_1$ related to the arbitrariness in the choice
of $\alpha$ in Eq. (\ref{EW41}). One can verify the commutation relations
(\ref{EW45}) are satisfied using that $2\sum_\mu F^*_\mu G_\mu=\langle
X_0|J|X_1\rangle = 1$.

%%%%%%%%%%%%%%%%%%%%%%%%%%%%%%%%%%%%%%%%%%%%%%%%%%%%%%%%%%%%%%%%%%%%%%%%%%%%%%%%%%%%%%%%%%

\section{Long range order, condensation and  the Bogoliubov spectrum of weakly interacting Bose gas}
\label{II3}

Consider an interacting  Bose gas trapped in an external  potential $V(\br)$ and
interacting with the two-body potential $U(|\br|)$. The Hamiltonian of the system
in the second quantization representation reads (see Eq. (\ref{EQ68}))
\[
H = \int\rd^3
\br\left\{\psi^\d(\br)\left(-\frac{\hbar^2}{2m}\nabla^2\right)
\psi(\br)+V(\br)\psi^\d(\br)\psi(\br)\right\}\qquad\qquad
\]
\be
+\frac12\int\rd^3\br\int\rd^3\br^\prime
U(|\br-\br^\prime|)\psi^\d(\br)\psi^\d(\br^\prime)\psi(\br^\prime)\psi(\br).
\en{U1}

According to the Penrose \& Onsager criterium \cite{PO}, there is a condensation in
the system of bosons,  related also to the concept of the long range order (see
also Ref. \cite{LRO}), if the one-particle statistical operator $\sigma_1$, defined
in Eq.~(\ref{sigmas}), has an eigenvalue on the order of the total number of
particles (we will use $\sigma_1$ instead of $\mathcal{R}_1$ since the approach is
applicable for the variable total number of particles and in our case it is not
fixed). Let us denote by $|\Phi_0\rangle$  the eigenstate of $\sigma_1$ with the
macroscopic eigenvalue $\langle n_0\rangle\sim \langle N\rangle$. It is a  state
from the  single-particle Hilbert space (the eigenvalue $\langle n_0\rangle$ is the
average population of this state, see also Eq. (\ref{EQ89}) of section
\ref{subI7B}). One could expand the operator $\sigma_1$ in the basis of its
eigenstates, however, below we will see that one must choose another basis for the
expansion in the complement of the Hilbert space orthogonal to the macroscopically
occupied  state $|\Phi_0\rangle$. For now we assume that the basis states
$|\Phi_\alpha\rangle$, $\alpha\ge0$ are taken from the single-particle Hilbert
space.   The statistical operator $\sigma_1$ assumes the following form (we assume
that the basis is countable and the eigenfunctions are integrable in the coordinate
representation)
\be
\sigma_1 = \langle a^\d_0a_0\rangle |\Phi_0\rangle\langle \Phi_0|+
\sum_{\alpha,\beta\ge1}\langle a^\d_\alpha a_\beta\rangle |\Phi_\beta\rangle\langle
\Phi_\alpha|\equiv \langle a^\d_0a_0\rangle |\Phi_0\rangle\langle
\Phi_0|+\sigma_1^{(\perp)}.
\en{U2}
The coefficients of this expansion are easily found by the observation that (see
Eq. (\ref{sigm}) of section \ref{subI7A})
\[
\langle\Phi_\beta|\sigma_1|\Phi_\alpha\rangle = \langle a^\d_\alpha a_\beta\rangle,
\]
where the boson operator $a_\alpha$ annihilates the state $|\Phi_\alpha\rangle$,
i.e. it appears in the expansion of the annihilation operator $\Psi(\br)$:
\be
\Psi(\br) = \sum_\alpha \Phi_\alpha(\br)a_\alpha.
\en{U3}

The  assumption of the long range order, i.e. $\langle n_0\rangle = \langle
a^\d_0a_0\rangle \sim \langle N\rangle$, is complemented that there is just one
macroscopically occupied state, i.e.   the eigenvalues $n^{(\perp)}_i$ of the
orthogonal complement $\sigma_1^{(\perp)}$  satisfy $n^{(\perp)}_i\ll \langle
n_0\rangle$. Then, for instance, the occupations of the states $\Phi_\alpha$,
$\alpha\ge 1$, are also small $\langle a^\d_\alpha a_\alpha\rangle \ll \langle
N\rangle$. The trace of $\sigma_1$ is equal to the average of the total number of
atoms, i.e. we have $\sum_\alpha\langle a^\d_\alpha a_\alpha\rangle = \langle N
\rangle$.

Following C.N.~Yang \cite{LRO}, one can easily show that in this case any
statistical operator $\sigma_s$ of the system of bosons  has an eigenvalue
$\sim\langle N\rangle^s$, for instance $\sigma_2$ has the largest eigenvalue which
is  bound as follows $\langle n_0\rangle(\langle n_0\rangle-1)\le \sigma_2\le
\langle N(N-1)\rangle$. This is clear from the following calculation $\langle
a^\d_0a^\d_0a_0a_0\rangle = \langle n_0^2\rangle - \langle n_0\rangle \ge \langle
n_0\rangle^2 - \langle n_0\rangle$, whereas the latter term is  one of the  the
diagonal terms of $\sigma_2$ in the above used basis and  is clearly less than the
maximal eigenvalue of $\sigma_2$. The statistical operators $\sigma_{1,2}$ are
those  used in the calculation of the average energy of the system, thus the long
range order allows to approximate  the expression for the system energy in the
orders of $\langle n_0\rangle$ (see below).

In the basis (\ref{U3}) the Hamiltonian reads
\be
H = \sum_{\alpha,\beta}\mathcal{E}_{\alpha,\beta}a^\d_\alpha a_\beta +
\frac12\sum_{{}^{\alpha_1,\alpha_2}_{\beta_1,\beta_2}}U^{\alpha_1,\alpha_2}_{\beta_1,\beta_2}a^\d_{\alpha_1}a^\d_{\alpha_2}
a_{\beta_2}a_{\beta_1},
\en{U4}
where the single-particle part of the Hamiltonian and the interaction part   have
the following expansion coefficients:
\be
\mathcal{E}_{\alpha,\beta} \equiv \int\rd^3
\br\left\{\Phi^*_\alpha(\br)\left(-\frac{\hbar^2}{2m}\nabla^2\right)\Phi_\beta(\br)+
\Phi^*_\alpha(\br)V(\br)\Phi_\beta(\br)\right\}
\en{DefE1}
\be
U^{\alpha_1,\alpha_2}_{\beta_1,\beta_2} \equiv \int\rd^3\br\int\rd^3\br^\prime
U(|\br-\br^\prime|)\Phi^*_{\alpha_1}(\br)
\Phi^*_{\alpha_2}(\br^\prime)\Phi_{\beta_2}(\br^\prime)\Phi_{\beta_1}(\br).
\en{DefU}
In the following we will  use the evident properties of these coefficients:
\be
\mathcal{E}_{\alpha,\beta} = (\mathcal{E}_{\beta,\alpha})^*,\quad
U^{\alpha_1,\alpha_2}_{\beta_1,\beta_2} =
(U^{\beta_1,\beta_2}_{\alpha_1,\alpha_2})^*,\quad
U^{\alpha_1,\alpha_2}_{\beta_1,\beta_2} = U^{\alpha_2,\alpha_1}_{\beta_2,\beta_1}.
\en{PropEU}

The   long range order assumption allows us to use the successive  approximations
of the system Hamiltonian (\ref{U4}) in the perturbation theory with the small
parameter $\langle n_0 \rangle^{-1/2}$. This assumption is sometimes related to the
appearance of a macroscopic order parameter and is related to the classical limit
of quantum description. However, one should yet supplement the long range order
condition by an additional one before one could talk about a macroscopic order
parameter. Namely,  that the distribution of the occupation number $n_0=a^\d_0a_0$
is peaked at the average value, i.e.
\be
\sqrt{\langle (n_0-\langle n_0\rangle)^2 \rangle} \ll \langle n_0\rangle.
\en{ASS}
We will also assume the practically relevant condition that the mean-field
potential created by the interaction between the bosons, i.e. the average
interaction potential, has the same order as the kinetic term:
\be
U^{0,0}_{0,0}\langle n_0\rangle \sim \mathcal{E}_{0,0}.
\en{WI}
Under this assumption, one can approximate the average value of  the Hamiltonian in
Eq. (\ref{U4}) in the leading order in $\langle n_0\rangle$ as follows
\[
\langle H \rangle = \mathcal{E}_{0,0}\langle n_0\rangle +
\frac{U^{0,0}_{0,0}}{2}\langle n_0(n_0-1)\rangle + \mathcal{O}(\sqrt{\langle
n_0\rangle})\approx \mathcal{E}_{0,0}\langle n_0\rangle +
\frac{U^{0,0}_{0,0}}{2}\langle n_0\rangle^2 +\mathcal{O}(\sqrt{\langle n_0\rangle})
\]
\be
 \equiv \langle H^{(0)}\rangle +
\mathcal{O}(\sqrt{\langle n_0\rangle}).
\en{U8}
We have denoted by $  H^{(0)}$ the leading term of the Hamiltonian on the  order
$\mathcal{O}(\langle n_0\rangle)$ \footnote{Here and below, the order of the
approximations to the Hamiltonian, the chemical potential and the free energy are
denoted by the superscript in brackets.   These correspond to the orders of
correction  these quantities describe  in the perturbation theory and do not
correspond to respective powers of $\langle n_0\rangle$ the approximations give.}.
It can be put in the form of the so-called Gross-Pitaevskii functional, i.e.
\[
 \langle H^{(0)}\rangle = E_{GP}(\Psi^*_0,\Psi_0) \equiv \int\rd^3 \br\left\{\Psi^*_0(\br)
\left(-\frac{\hbar^2}{2m}\nabla^2\right)\Psi_0 + V(\br)|\Psi_0(\br)|^2\right\}
\]
\be
+\frac{1}{2} \int\rd^3\br\int\rd^3\br^\prime
U(|\br-\br^\prime|)|\Psi_0(\br)|^2|\Psi_0(\br^\prime)|^2,
\en{U9}
by introducing the order parameter $\Psi_0(\br) = \langle n_0\rangle\Phi_0(\br)$ of
the condensate \cite{PSBook}.

To find the equation defining the order parameter, consider the   ground state of
the system, which is the extremal of the average energy
\be
\bar{E}(| G\rangle,\langle G|) \equiv \frac{\langle G|H| G\rangle}{\langle G|
G\rangle},
\en{U5}
where $| G\rangle$ is the tentative system state. Indeed, variation of $\bar{E}(|
G\rangle,\langle G|)$ with respect to the bra-vector $\langle  G|$ leads to the
Euler-Lagrange equation $H| G\rangle = \bar{E}(| G\rangle,\langle G|)| G\rangle$.
In our case it should be taken under the condition that the average number of
bosons $\langle N\rangle $ is fixed. To account for the latter condition, one can
use the  functional
\be
\bar{F}(| G\rangle,\langle G|) = \bar{E}(| G\rangle,\langle G|) - \mu \frac{\langle
G|N| G\rangle}{\langle G| G\rangle}
\en{U6}
with the Lagrange multiplier $\mu$. The purpose of the functional $\bar{F}$ is to
convert the constrained minimization problem into an unconstrained one, the $\mu$
is defined by equating the derivative over $\langle N\rangle$ to zero: $\frac{\p
\bar{F}}{\p \langle N\rangle} = 0$, thus  we have $\mu = \frac{\p \bar{E}}{\p
\langle N\rangle}$.

The order parameter or the ``condensate wave function'' $\Psi_0(\br)$ can be found
by minimizing the functional $\bar{F}$ (\ref{U6}), which in the made approximation
($N \approx n_0$) becomes
\be
F_{GP} \equiv E_{GP}(\Psi^*_0,\Psi_0)- \mu^{(0)} \int\rd^3 \br|\Psi_0(\br)|^2,
\en{FGP}
where that the Lagrange multiplier $\mu^{(0)}$ now accounts for the normalization
of the order parameter: $ \int\rd^3 \br|\Psi_0(\br)|^2 = \langle n_0\rangle$.
Taking the first variation in $\Psi^*_0(\br)$ and setting $\Psi_0(\br) =\langle
n_0\rangle \Phi_0(\br)$ we obtain the Euler-Lagrange equation for the strongly
occupied  state
\be
\mu^{(0)} \Phi_0(\br) = -\frac{\hbar^2}{2m}\nabla^2\Phi_0(\br) + \langle n_0\rangle
\left\{\int\rd^3\br^\prime U(|\br-\br^\prime|)|\Phi_0(\br^\prime)|^2
\right\}\Phi_0(\br).
\en{U11}
Multiplying this equation by $\Phi^*(\br)$ and integrating we obtain the chemical
potential
\be
\mu^{(0)} = \mathcal{E}_{0,0} + \langle n_0\rangle U^{0,0}_{0,0}.
\en{mu}

The next order approximation  to the energy of  a gas in an external potential will
be considered below after we consider a much  simpler system -- the weakly
interacting Bose gas in a box, whose excitation spectrum was considered by N.N.
Bogoliubov \cite{BgMeth}.

%%%%%%%%%%%%%%%%%%%%%%%%%%%%%%%%%%%%%%%%%%%%%%%%%%%%%%%%%%%%%%%%%%%%%%%%%%%%%%%
\subsection{The excitation spectrum of weakly interacting Bose gas in a box}
\label{subII3A}

Consider the weakly interacting Bose gas in free space. We will work in a three
dimensional box $V = L^3$ of the size $L$, and consider the limit $L\to \infty$. By
weakly we mean that the average potential created due to the interaction between
the bosons is negligible compared to the kinetic energy of the bosons. Then the
strongly occupied state in this case is the zero-momentum state $\bp = 0$ (which is
approximately the ground state of the system in our case). Thus we will use the
momentum states for the expansion, i.e. $\Phi_\alpha(\br) =
\frac{1}{\sqrt{V}}e^{i\bp_\alpha \br/\hbar}$, where $\bp_\alpha = \frac{2\pi
\mathbf{n}_\alpha}{L}$ with $\mathbf{n}_\alpha$ being three dimensional vector
taking integer values. The Hamiltonian (\ref{U1}) in this case has the following
form (we use the summation over the discrete momenta $\bp$)
\be
H = \sum_\bp \frac{\bp^2}{2m} a^\d_\bp a_\bp + \frac12
\sum_{{}^{\bp_1,\bp_2}_{\bp^\prime_1,\bp^\prime_2} }
\frac{\hat{U}(\bp_1-\bp^\prime_1)}{V}\delta_{\bp_1+\bp_2,\bp^\prime_1+\bp^\prime_2}a^\d_{\bp_1
} a^\d_{\bp_2} a_{\bp^\prime_2} a_{\bp^\prime_1},
\en{Hbox}
where $\hat{U}$ is  the Fourier transform of the interaction potential,
\be
\hat{U}(\bp) = \int \rd^3\br\, e^{i\frac{\bp \br}{\hbar}}U(|\br|) = 4\pi\hbar
\int\rd r r^2 U(r)\frac{\sin\left(\frac{pr}{\hbar}\right)}{ pr },
\en{hatU}
which is obviously a function of $p^2 = \bp^2$. For the diagonalization of the
quadratic form (the form $F^{(2)}$ (\ref{F2}) below) to be possible one must demand
that for all $\bp\ne0$ the Fourier transform of the interaction is non-negative
$\hat{U}(\bp)\ge0$. This condition is satisfied, for instance, by the zero-range
repulsive interactions $U(\br) = g\delta(\br)$ with $g>0$.

The leading order part of the Hamiltonian (\ref{Hbox})  is
\be
H^{(0)} = \frac12 \frac{\hat{U}(0)}{V}n_0(n_0-1),
\en{H0}
thus using the unconstrained minimization problem for  $\langle F^{(0)}\rangle =
\langle H^{(0)}\rangle - \mu^{(0)}\langle n_0 \rangle $ and the assumption
(\ref{ASS}) we get approximately $\mu^{(0)} = \frac{\hat{U}(0)}{V}\langle n_0
\rangle$. Taking into account the momentum conservation (given by the delta-symbol
in Eq. (\ref{Hbox})) we get the next order of $H$ as follows (note that this term
is actually a forth order term in the boson operators and \textit{preserves} the
total number of bosons in the system)
\[
H^{(2)} = \sum_{\bp\ne0} \biggl\{\left[\frac{\bp^2}{2m}  +
\left(\frac{\hat{U}(0)}{V}+\frac{\hat{U}(\bp)}{V}\right)n_0\right]a^\d_\bp a_\bp
\qquad \qquad
\]
\be
+\frac12 \frac{\hat{U}(\bp)}{V} \left[(a^\d_0)^2 a_\bp a_{-\bp} +  (a_0)^2 a^\d_\bp a^\d_{-\bp}
\right]\biggr\}.
\en{H2}
Below, we work with the free energy operator $F = H - \mu N$ instead of the
Hamiltonian. We also note that, due to the momentum conservation and weak
interaction, the expansion of the free energy contains  no first order term, i.e.
\[
F = F^{(0)}+ F^{(2)}+\ldots = H^{(0)} - \mu^{(0)}n_0 + H^{(2)} -\mu^{(0)} N^{(2)}
-\mu^{(2)} n_0 + \ldots,
\]
where $N^{(2)} = \sum_{\bp\ne0}a^\d_\bp a_\bp$.

In the next order approximation to the Hamiltonian  (more precisely,  to  the
average energy functional (\ref{U5})) is given by the terms quadratic in the
creation and annihilation operators $a^\d_\bp$ and $a_\bp$ for $\bp\ne0$ due to the
momentum conservation by the interaction term. To see how the averages involving
$a^\d_0$ or $a_0$   can be approximated in our approach we introduce the polar
decomposition of the operators $a^\d_0$ and $a_0$:
\be
a^\d_0 = e^{i\varphi_0}\sqrt{n_0+1} = \sqrt{n_0}e^{i\varphi_0},\quad a_0 =
e^{-i\varphi_0}\sqrt{n_0} = \sqrt{n_0+1}e^{-i\varphi_0},
\en{U12}
where the shift operator $e^{i\varphi_0}$  acts in the following way on the Fock
state
\be
e^{i\varphi_0}|n_0\rangle = |n_0+1\rangle.
\en{U13}
The following immediate identities are found by application of Eq. (\ref{U12})
\be
e^{-i\varphi_0}e^{i\varphi_0} = I - |N_0\rangle\langle  N_0|, \quad
e^{i\varphi_0}e^{-i\varphi_0} = I -|0\rangle\langle 0|,
\en{U14}
where $|N_0\rangle$ is the state with the maximal occupation number, if it exists.
Thus, strictly speaking, the shift operator is not unitary (hence, the ``phase''
$\varphi_0$ is not Hermitian operator). However,  in our case we can safely
approximate $e^{i\varphi_0}e^{-i\varphi_0} = e^{-i\varphi_0}e^{-i\varphi_0} =  I$
neglecting the states of $n_0=0$ and $n_0 = N_0$ (the latter state corresponds to
all atoms occupying the state $\Phi_0(\br)$) which are the boundary states and are
not important. In this approximation, we can perform the transformation
\be
\hat{a}_\bp =e^{i\varphi_0}{a}_\bp, \quad \hat{a}^\d_\bp =
e^{-i\varphi_0}{a}^\d_\bp, \quad  \bp\ne0,
\en{U15}
which is obviously  canonical one (note that the shift operator commutes with the
orthogonal degrees of freedom described by the operators $a_\bp$ with $\bp\ne0$).
In this respect  we should mention the recent   results on the validity of the
$U(1)$-symmetry breaking approach,  where it is argued that one can substitute the
operators $a_0$ and $a^\d_0$ by the classical variables in the Hamiltonian
augmented by a symmetry-breaking term   \cite{BgAppr}.

Using  the canonical transformation (\ref{U15}), the identities  $a_0^2 =\break
e^{-2i\varphi_0}\sqrt{n_0(n_0-1)}$ and (\ref{U14}), replacing the operator $n_0$ by
its average value due to the assumption (\ref{ASS}), we get the second order
correction to the    free energy $F$, i.e. $F^{(2)} = H^{(2)} -\mu^{(0)} N^{(2)} $
(where the scalar term $-\mu^{(2)}\langle n_0\rangle$ is inessential and dropped),
as follows
\be
F^{(2)}  = \sum_{\bp\ne0} \left\{\left[\frac{\bp^2}{2m}  + \frac{\hat{U}(\bp)}{V}
\langle n_0\rangle \right] \hat{a}^\d_\bp \hat{a}_\bp +\frac12
\frac{\hat{U}(\bp)}{V} \left[\langle n_0\rangle^2 \hat{a}_\bp \hat{a}_{-\bp} +
\langle n_0\rangle^2 \hat{a}^\d_\bp \hat{a}^\d_{-\bp} \right]\right\}.
\en{F2}
Note that $F^{(2)}$ consists of block operators for each value of  $\bp$. Each such
block can be diagonalized by the Bogoliubov transformation which involves just two
modes with the opposite momenta: $\bp$  and $-\bp$. Denoting
\be
\alpha(p^2) =  \frac{\bp^2}{2m} + \frac{\hat{U}(\bp)}{V} \langle n_0\rangle, \quad
\beta(p^2) = \frac{\hat{U}(\bp)}{V}\langle n_0\rangle^2
\en{alphbet}
we can  write  $F^{(2)} = \sum_\bp F^{(2)}_\bp$  where each  $\bp$-block is in the
standard form  (note that we use two term block: the term with $\bp$ and with
$-\bp$ enter each block, hence the free term is twice the free term for $\bp$)
\be
F^{(2)}_\bp  = \frac12(\hat{a}^\d_\bp, \hat{a}^\d_{-\bp},
\hat{a}_\bp,\hat{a}_{-\bp})\left(\begin{array}{cc} A_\bp & B_\bp \\ B_\bp &
A_\bp\end{array}\right)\left(\begin{array}{l}\hat{a}_\bp \\ \hat{a}_{-\bp} \\
\hat{a}^\d_\bp \\ \hat{a}^\d_{-\bp} \end{array}\right) -\alpha(p^2),
\en{F2p}
where
\[
A_\bp = \left(\begin{array}{cc} \alpha(p^2) & 0 \\0 & \alpha(p^2)\end{array}\right),
\quad B_\bp = \left(\begin{array}{cc} 0 & \beta(p^2) \\ \beta(p^2) & 0\end{array}\right),
\]
note that both $A_\bp$  and $B_\bp $  are real. According to  the general theory of
section \ref{secII2} the diagonalization of the quadratic form given in Eq.
(\ref{F2p}) is performed by the transformation composed of the eigenvectors of the
matrix $J \mathcal{H}$ with  the Hessian defined in Eq. (\ref{F2p}) by the matrices
$A_\bp$ and $B_\bp$. We obtain the characteristic polynomial $\chi(\lambda) =
\mathrm{det}(J\mathcal{H} - \lambda I)$ in the form $\chi(\lambda) = (\lambda^2
+\beta^2 - \alpha^2)^2$ thus the eigenvalues $\lambda_\pm = \pm\sqrt{\alpha^2 -
\beta^2}$ are doubly degenerate. The eigenvectors $|\lambda\rangle$ can be easily
found, the right normalization eigenvectors, i.e. satisfying $\langle
\lambda|J|\lambda\rangle =1$ correspond to the positive eigenvalues $\lambda_+$. We
have the transformation matrix $S$ in the form
\be
S =  \left(\begin{array}{cc} U & V\\ V & U\end{array}\right)
\en{Sp}
where
\[
U = \frac{1}{\sqrt{2}(\alpha^2-\beta^2)^{\frac14}}\left(\begin{array}{cc}
\frac{\beta}{\left[\alpha - \sqrt{\alpha^2-\beta^2}\right]^{\frac12}} & 0 \\
 0 & \frac{\beta}{\left[\alpha - \sqrt{\alpha^2-\beta^2}\right]^{\frac12}} \end{array}\right),
\]
\[
 V  = \frac{1}{\sqrt{2}(\alpha^2-\beta^2)^{\frac14}}\left(\begin{array}{cc}
0& -\frac{\beta}{\left[\alpha + \sqrt{\alpha^2-\beta^2}\right]^{\frac12}}  \\
 -\frac{\beta}{\left[\alpha + \sqrt{\alpha^2-\beta^2}\right]^{\frac12}} & 0
\end{array}\right).
\]
For instance, we obtain the excitation spectrum, the Bogoliubov spectrum as follows
\be
E(p^2) = \lambda_+  = \sqrt{\alpha^2 -\beta^2 } = \sqrt{\frac{\bp^2}{2m} \left(
\frac{\bp^2}{2m} + \frac{2\hat{U}(\bp)}{V}\langle n_0\rangle^2\right)}.
\en{Ep}
The Bogoliubov transformation (\ref{EW24}) makes the operator $F^{(2)}$ diagonal  \
\be
F^{(2)} = \sum_{\bp\ne 0} E(p^2) b^\d_{\bp}b_{\bp}- \sum_{p^2}(E(p^2) - \alpha(p^2)).
\en{F2pd}

%%%%%%%%%%%%%%%%%%%%%%%%%%%%%%%%%%%%%%%%%%%%%%%%%%%%%%%%%%%%%%%%%%%%%%%%%%%%%%%%%%%%%%%%%%%%%%

\subsection{The excitation spectrum of  an interacting Bose gas in an external potential}
\label{subII3B}

%%%%%%%%%%%%%%%%%%%%%%%%%%% FIRST ORDER CORRECTION %%%%%%%%%%%%%%%%%%%%%%%%%%%%%%%%%%

Let us now return to the general case of a Bose gas in an external potential, which
we have started to  consider in section \ref{II3}. The interaction is not assumed
to be negligible as for the gas in a box considered above, but on the order of the
single-particle energy term, as in Eq. (\ref{WI}). We will consider the next two
terms  of the approximation in the orders of $\langle n_0\rangle$ of the
Hamiltonian (\ref{U4}). First of all, similar as in the case of the gas in a box,
the condition of the long range order, supplemented  by the conditions (\ref{ASS})
and (\ref{WI}), allows us to separate the amplitude and phase of the creation and
annihilation operators $a_0=e^{-i\varphi_0}\sqrt{n_0}$ and $a^\d_0=
\sqrt{n_0}e^{i\varphi_0}$,   as in Eq. (\ref{U12}). Thus, similar as   in the case
of a gas in a box, Eq. (\ref{U15}),  we perform a canonical transformation on the
rest of the boson operators, corresponding to the orthogonal modes:
\be
\hat{a}_\alpha =  e^{i\varphi_0}{a}_\alpha, \quad \hat{a}^\d_\alpha =
e^{-i\varphi_0}{a}^\d_\alpha, \quad  \alpha\ge1.
\en{U81}
The next order approximation to the Hamiltonian (\ref{U4}), which follows
$\mathcal{E}$ (\ref{U9}) in the expansion in the orders of $\langle n_0\rangle$,
is linear in the creation and annihilation operators $a^\d_\alpha$ and $a_\alpha$,
for $\alpha\ge1$:
\[
H^{(1)} = \sum_{\alpha\ge1}\left\{\left[\mathcal{E}_{\alpha,0}a_0 +
U^{\alpha,0}_{0,0}a^\d_0 a_0^2 \right]a^\d_\alpha +
\left[\mathcal{E}_{0,\alpha}a^\d_0 + U^{0,0}_{\alpha,0}(a^\d_0)^2 a_0
\right]a_\alpha\right\}
\]
\be
=\sum_{\alpha\ge1}\left\{ C_\alpha\hat{a}^\d_\alpha + C^*_\alpha
\hat{a}_\alpha\right\},
\en{U82}
where we have defined the  coefficients
\be
C_\alpha = \left[\mathcal{E}_{\alpha,0} + U^{\alpha,0}_{0,0}n_0\right]\sqrt{n_0}
\en{U83}
and in the derivation used the    properties (\ref{PropEU}). The conditions
(\ref{ASS}) and (\ref{WI}) allow us to replace $n_0$ in the definition of
$C_\alpha$ by the scalar $\langle n_0\rangle$ which substitution turns the
coefficient into a scalar. The definition of this scalar coefficient has the form
of a ``matrix element''  of the average boson energy in the macroscopically
occupied state, i.e. the Bose-Einstein condensate, between the two single particle
states $|\Phi_0\rangle $ and $|\Phi_\alpha\rangle$ for $\alpha \ge 1$ (see the
definitions (\ref{DefE1}) and (\ref{DefU})). Here the operator $\mathcal{E}$ is the
single-particle energy
\be
\mathcal{E} \equiv -\frac{\hbar^2}{2m}\nabla^2 + V(\br),
\en{U85}
while  the condensate energy also involves the average interaction potential
defined as
\be
\bar{U}_0(\br) \equiv \langle n_0\rangle \int\rd^3\br^\prime\,
U(|\br-\br^\prime|)|\Phi_0(\br^\prime)|^2 = \int\rd^3\br^\prime\,
U(|\br-\br^\prime|)|\Psi_0(\br^\prime)|^2.
\en{U86}
But then, using the equation (\ref{U11}), which  the order parameter
$|\Phi_0\rangle$ satisfies, one immediately concludes that $C_\alpha=0$:
\be
C_\alpha =  \sqrt{\langle n_0\rangle}\langle\Phi_\alpha|\mathcal{E} +
\bar{U}_0|\Phi_0\rangle = \mu^{(0)}\sqrt{\langle
n_0\rangle}\langle\Phi_\alpha|\Phi_0\rangle = 0,\quad \alpha\ge 1.
\en{U84}
Therefore, $C_\alpha=0$ for  the  arbitrary   basis $|\Phi_\alpha\rangle$,
$\alpha\ge1$, from the complement of the single-particle Hilbert space  orthogonal
to $|\Phi_0\rangle$. Note that the corresponding correction to the free energy
$F^{(1)} = H^{(1)}=0$, since the term $-\mu^{(0)} N^{(1)} - \mu^{(1)} n_0$ is also
identically zero ($\mu^{(1)} = \frac{\p \langle H^{(1)}\rangle }{\p \langle
n_0\rangle}=0$ and  $N^{(1)} = 0$).

%%%%%%%%%%%%%%%%%%%%%%%%%%% SECOND ORDER CORRECTION %%%%%%%%%%%%%%%%%%%%%%%%%%%%%%%%%%

We see that the first non-zero correction after $H^{(0)}$ to the Hamiltonian
(\ref{U4}), and, hence, to the free energy $F$,  is quadratic in the operators
$a^\d_\alpha$ and $a_\alpha$, $\alpha\ge1$. We obtain this quadratic term in the
following form (note that the term $F^{(2)}$ is forth order in the boson operators
and \textit{preserves} the total number of bosons in the system, as it should be)
\[
F^{(2)} = H^{(2)} - \mu^{(0)} N^{(2)} =
\sum_{\alpha,\beta\ge1}\left[\mathcal{E}_{\alpha,\beta} +
\left(U^{\alpha,0}_{\beta,0}+U^{\alpha,0}_{0,\beta}\right)n_0
-\mu^{(0)}\delta_{\alpha,\beta} \right] a^\d_\alpha a_\beta \qquad\qquad
\]
\be
+ \frac12\sum_{\alpha,\beta\ge1}U^{\alpha,\beta}_{0,0}a_0^2a^\d_\alpha a^\d_\beta +
\frac12\sum_{\alpha,\beta\ge1}U^{0,0}_{\alpha,\beta}(a^\d_0)^2a_\alpha a_\beta,
\en{U88}
where  $\mu^{(0)}$ is given by Eq. (\ref{mu})  and to derive this expression we
have used once again the properties (\ref{PropEU}) (we also have used the  term
$\mu^{(2)}n_0$ is approximately scalar due to the assumption (\ref{ASS}) and can be
safely  dropped). By performing the canonical transformation (\ref{U81}) in the
Hamiltonian (\ref{U88}) and replacing the operator $n_0$ by its average value, due
to the condition (\ref{ASS}), we get
\be
F^{(2)} = \sum_{\alpha,\beta\ge1}A_{\alpha,\beta} \hat{a}^\d_\alpha \hat{a}_\beta +
\frac12\sum_{\alpha,\beta\ge1}B^*_{\alpha,\beta}\hat{a}^\d_\alpha \hat{a}^\d_\beta
+ \frac12\sum_{\alpha,\beta\ge1}B_{\alpha,\beta}\hat{a}_\alpha \hat{a}_\beta,
\en{U89}
where we have introduced the matrices $A$ and $B$ with the matrix elements (with
$\alpha\ge1$ and $\beta \ge1$)
\be
A_{\alpha,\beta} = \mathcal{E}_{\alpha,\beta} +
\left(U^{\alpha,0}_{\beta,0}+U^{\alpha,0}_{0,\beta}\right)\langle n_0\rangle
-\mu^{(0)}\delta_{\alpha,\beta},
\en{A}
\be
\quad B_{\alpha,\beta} =U^{0,0}_{\alpha,\beta}\langle n_0\rangle
\en{B}
and  the obvious properties $A^\d = A$ and $\widetilde{B} = B$. The quadratic form
(\ref{U89}) is in the standard notations of Eq. (\ref{EW29}) of section
\ref{secII2}. Note   the physical meaning of the interaction potential terms in the
expression for $A$:
\[
U^{\alpha,0}_{\beta,0}\langle n_0\rangle = \langle n_0\rangle
\langle\Phi^{(1)}_0(\br^\prime)|\langle\Phi^{(2)}_\alpha(\br)|
U(|\br-\br^\prime|)|\Phi^{(1)}_0(\br^\prime)\rangle|\Phi^{(2)}_\beta(\br)\rangle
\]
\be
= \langle\Phi_\alpha(\br)|\bar{U}(\br)|\Phi_\beta(\br)\rangle,
\en{U91}
which is the matrix element of the average interaction potential  created by the
bosons in  the condensate taken  between the excitation  basis states   and
\be
U^{\alpha,0}_{0,\beta}\langle n_0\rangle = \langle
n_0\rangle\langle\Phi^{(1)}_0(\br^\prime)|\langle\Phi^{(2)}_\alpha(\br)|
U(|\br-\br^\prime|)|\Phi^{(1)}_\beta(\br^\prime)\rangle|\Phi^{(2)}_0(\br)\rangle,
\en{U92}
which has the meaning of the exchange interaction between the condensed  and
non-condensed  bosons,  corresponding to the matrix element (\ref{U91}). Precisely
these two terms appear in the classical linearization of the Gross-Pitaevskii
energy functional about the order parameter $\Psi_0(\br)$ with the only difference
that the linearization is considered for the variations lying in the orthogonal
complement to the order parameter itself. Indeed, consider the classical free
energy functional (\ref{FGP}) about the order parameter of the condensate, which is
i.e. by Eq. (\ref{U11}) its extremal, $\frac{\delta F_{GP}}{\delta \Psi^*_0(\br)} =
0$. To simplify considerably the notations, consider only  the case  when
$U(|\br-\br^\prime|) = g\delta(\br-\br^\prime)$. We have the second-order
correction at the order parameter $\Psi_0(\br)$ as follows
\be
F^{(2)}_{GP} = \int\rd^3\br(\delta \Psi^*(\br),\delta
\Psi(\br))\left[\mathcal{H}_{GP}(\br) - \mu \right] \left(\begin{array}{c} \delta \Psi(\br) \\
\delta \Psi^*(\br) \end{array} \right),
\en{F2GP}
where the Hessian reads
\be
\mathcal{H}_{GP}(\br) \equiv \left(\begin{array}{cc} \mathcal{E} +
2g|\Psi_0(\br)|^2   & g \Psi^2_0(\br)
\\ g [\Psi^*(\br)]^2 & \mathcal{E} + 2g|\Psi_0(\br)|^2  \end{array}\right),
\en{HessGP}
with $\mathcal{E}$ given by   Eq. (\ref{U85}). Comparing the expressions
(\ref{F2GP}) and (\ref{U89}) with the use of the expansion (\ref{U3}) and the
transformation (\ref{U81}) we observe that the quantum quadratic form $F^{(2)}$
(\ref{U89})  can be put as follows
\be
F^{(2)} =\int\rd^3\br(\delta \Psi_\perp^\d(\br),\delta
\Psi_\perp(\br))\left[\mathcal{H}_{GP}(\br) - \mu \right] \left(\begin{array}{c} \delta \Psi_\perp(\br) \\
\delta \Psi^\d_\perp(\br) \end{array} \right),
\en{F2HGP}
with
\be
\delta \Psi_\perp(\br) \equiv \sum_{\alpha\ge1}\hat{a}_\alpha \Phi(\br).
\en{dPsi_perp}
In this respect, the above  approach has similarities with   the approach  by
Y.~Castin \& R.~Dum \cite{CD-BEC}, where a systematic expansion in the fraction of
the non-condensed atoms is considered without the use of the symmetry breaking. In
our case, the  quadratic form in $F^{(2)}$ is also equivalent to the classical
quadratic form obtained from the second-order correction to the  classical free
energy functional (\ref{FGP})  at the order parameter if the latter is considered
in the linear space of all variations  orthogonal to the order parameter itself.

In our approach the zero mode corresponding to the usual $U(1)$-symmetry breaking
in the Bogoliubov-de Gennes approach (see for instance, Ref. \cite{NPBook}) does
not appear, since we treat the phase of the condensate as a quantum variable with
an undefined value. For instance, in the extreme case,  the system  can be
considered to be in the combination of the Fock states
\[
|G\rangle = \sum_{\{n_\alpha\ll\langle N\rangle
\},\alpha\ge1}C(\{n_\alpha\})|n_0,\{n_\alpha\}\rangle,
\]
which satisfies the long range order assumption for $n_0\sim\langle N\rangle$ and
condition (\ref{ASS})). Instead, we use the conditions of the long range order and
(\ref{ASS}) which are enough to assume the classical limit, while the phase,
namely,  the exponential operator $e^{i\varphi_0}$  is incorporated into the new
operators $\hat{a}_\alpha$ by a canonical transformation (\ref{U81})).

%%%%%%%%%%%%%%%%%%%%%%%%%%%%%%%%%%%%%%%%%%%%%%%%%%%%%%%%%%%%%%%%%%%%%%%%%%%%%%%%%%%%%%%%%%%%%
%%%%%%%%%%%%%%%%%%%%%%%%% APPLICATION TO SPIN SYSTEMS %%%%%%%%%%%%%%%%%%%%%%%%%%%%%%%%%%%%%%%
%%%%%%%%%%%%%%%%%%%%%%%%%%%%%%%%%%%%%%%%%%%%%%%%%%%%%%%%%%%%%%%%%%%%%%%%%%%%%%%%%%%%%%%%%%%%%%

\section{The Jordan-Wigner transformation:  fermionization of interacting 1D spin chains}
\label{secII4}

In chapter \ref{chI} we have seen that  the creation/annihilation  operator can be
defined as  some linear transformation of the states of a system of
indistinguishable particles which increases/decreases the total number of particles
in the system and that  their commutation relations can be \textit{derived}.
However, the commutation relations, that a set of the creation and annihilation
operators satisfy, themselves  define all other properties of these operators,
including the structure of the associated Hilbert space, e.g. the space dimension
and the Fock basis. Any system which can be described by a set of operators which
satisfy the same commutation relations is thus equivalent to a system of
indistinguishable particles. This is the idea of the Jordan-Wigner transformation
for   fermionization of the spin models.

Consider the set of operators $a_j$, $j=1,\ldots,s$ and their Hermitian conjugates
which satisfy the anti-commutation relations
\be
\{a_j,a_k\} = 0, \quad \{a_j,a^\d_k\} = \delta_{j,k}.
\en{EJ1}
From these it follows that $a^2_j=0$ and $(a^\d_j)^2 = 0$. Define the Hermitian
operators $\hat{n}_j = a^\d_j a_j$. We have from Eq. (\ref{EJ1})
\be
\hat{n}_j^2 =a^\d_j a_j a^\d_j a_j = a^\d_j (-a^\d_ja_j +1)a_j = a^\d_j a_j =
\hat{n}_j,
\en{EJ2}
\be
\hat{n}_j\hat{n}_k = a^\d_j a_ja^\d_k a_k =  a^\d_j (-a^\d_k a_j)a_k =  a^\d_j
a^\d_k a_ka_j = ... = \hat{n}_k \hat{n}_j, \quad k\ne j,
\en{EJ3}
i.e. the set of the observables $\hat{n}_j$ has the common basis, where each has
just two eigenvalues: $0$ or $1$. The eigenstates can be labelled by the occupation
numbers of the operators $\hat{n}_j$ (which we will denote by  $n_j$)
$\vr{n_1,n_2,\ldots,n_s}$, where $n_j\in\{0;1\}$. Moreover, from the
anti-commutators (\ref{EJ1}) we obtain
\be
[\hat{n}_j,a_k] = a^\d_j a_j a_k - a_k a^\d_j a_j = -\delta_{j,k}a_j,\quad
[\hat{n}_j,a^\d_k] = \delta_{j,k}a^\d_j.
\en{EJ4}
Hence,
\[
\hat{n}_j a_j\vr{n_1,\ldots,n_s }= (n_j-1)a_j\vr{n_1,\ldots,n_s }=0,
\]
\[
\hat{n}_j a^\d_j\vr{n_1,\ldots,n_s }= (n_j+1)a^\d_j\vr{n_1,\ldots,n_s
}=a^\d_j\vr{n_1,\ldots,n_s }, \quad
\]
(in the first case $n_j=1$ and in the second case  $n_j=0$, otherwise the vector
itself  must be zero) where the non-zero state vectors on the l.h.s. are
normalized, since
\[
\vl{n_1,\ldots,n_s}a^\d_ja_j\vr{n_1,\ldots,n_s } = n_j,\quad
\vl{n_1,\ldots,n_s}a_ja^\d_j\vr{n_1,\ldots,n_s } = n_j+1.
\]
Therefore, the action of the operators $a^\d_j$ and $a_j$ is similar to the action
of the fermionic creation and annihilation operators on the Fock states:
\be
a_j\vr{n_1,\ldots,n_j,\ldots,n_s } =
e^{i\varphi^{(-)}_j}\vr{n_1,\ldots,n_j-1,\ldots,n_s },
\en{EJ5}
\be
a^\d_j\vr{n_1,\ldots,n_j,\ldots,n_s } =
e^{i\varphi^{(+)}_j}\vr{n_1,\ldots,n_j+1,\ldots,n_s },
\en{EJ6}
where the phases $\varphi^{(\pm)}_j$ depend on  the ordering of the occupation
numbers of the Fock basis states. Denoting the state $\vr{0,\ldots,0} \equiv
\vr{Vac}$ we can set, for instance,
\be
\vr{n_1,\ldots,n_s } \equiv
(a^\d_1)^{n_1}(a^\d_2)^{n_2}\cdots(a^\d_s)^{n_s}\vr{Vac},
\en{EJ7}
then due to the anti-commutation relations (\ref{EJ1}) Eqs. (\ref{EJ5}) and
(\ref{EJ6}) become
\be
a^\d_j\vr{n_1,\ldots,n_j,\ldots,n_s}
=(-1)^{\sum_{k=1}^{j-1}n_k}\vr{n_1,\ldots,n_j+1,\ldots,n_s},
\en{EJ8}
\be
a_j\vr{n_1,\ldots,n_j,\ldots,n_s}
=(-1)^{\sum_{k=1}^{j-1}n_k}\vr{n_1,\ldots,n_j-1,\ldots,n_s},
\en{EJ9}
where in Eq. (\ref{EJ8}) $n_j=0$ and in Eq. (\ref{EJ9}) $n_j=1$, otherwise the
r.h.s. is zero. Indeed, we have
\[
a^\d_j\vr{n_1,\ldots,n_j,\ldots,n_s} =
a^\d_j(a^\d_1)^{n_1}\cdots(a^\d_s)^{n_s}\vr{Vac}
\]
\[
= (-1)^{\sum_{k=1}^{j-1}n_k}(a^\d_1)^{n_1}\cdots
(a^\d_j)^{n_j+1}\cdots(a^\d_s)^{n_s}\vr{Vac}
\]
\[
=\delta_{n_j,0}(-1)^{\sum_{k=1}^{j-1}n_k}\vr{n_1,\ldots,n_j+1,\ldots,n_s},
\]
%%%%%%%%%%%%%%%%%%%%%%%%%%%%%%%%%%%%%%%%%%%%%%%%%%%%%%%%%%%%%%%%%%%%%%%%
\[
a_j\vr{n_1,\ldots,n_j,\ldots,n_s} = a_j(a^\d_1)^{n_1}\cdots(a^\d_s)^{n_s}\vr{Vac}
\]
\[
= (-1)^{\sum_{k=1}^{j-1}n_k}(a^\d_1)^{n_1}\cdots
a_j(a^\d_j)^{n_j}\cdots(a^\d_s)^{n_s}\vr{Vac}
\]
\[
=(-1)^{\sum_{k=1}^{s}n_k}(a^\d_1)^{n_1}\cdots \cdots(a^\d_s)^{n_s}a_j\vr{Vac}
\]
\[
+ \delta_{n_j,1}(-1)^{\sum_{k=1}^{j-1}n_k}\vr{n_1,\ldots,n_j-1,\ldots,n_s}
\]
\[
=\delta_{n_j,1}(-1)^{\sum_{k=1}^{j-1}n_k}\vr{n_1,\ldots,n_j-1,\ldots,n_s}.
\]

The Hilbert space $\mathcal{H}$ can in fact have several different vacuum states
$\vr{Vac,\mu}$ which may depend on additional observables $\mu$. For each fixed set
of observables  $\mu$ there is the Fock basis constructed above.  Thus, in general,
the Hilbert space  can be represented as a tensor product of the Fock space and an
independent space pertaining to the observables $\mu$.

Consider now a single spin $\frac12$. We have the Pauli matrices
\be
\sigma_x = \left(\begin{array}{cc}0 & 1 \\ 1 & 0 \end{array} \right), \quad
\sigma_y = \left(\begin{array}{cc}0 & -i \\ i & 0  \end{array}\right), \quad
\sigma_z = \left(\begin{array}{cc}1 & 0 \\ 0 & -1  \end{array}\right).
\en{EJ10}
The eigenstates of the $\sigma_z$ are $\vr{z\uparrow}$ and $\vr{z\downarrow}$ with
the eigenvalues $\pm1$. Denote the raising and lowering matrices $\sigma_\pm$,
where
\be
\sigma_+ = \left(\begin{array}{cc}0 & 1 \\ 0 & 0 \end{array}
\right)=\frac{\sigma_x+i\sigma_y}{2}, \quad\sigma_- = \left(\begin{array}{cc}0 & 1 \\
0 & 0\end{array} \right)=\frac{\sigma_x-i\sigma_y}{2},
\en{EJ11}
so that $\sigma_+\vr{z\uparrow} = 0$, $\sigma_+\vr{z\downarrow} = \vr{z\uparrow}$,
$\sigma_-\vr{z\uparrow} = \vr{z\downarrow}$, $\sigma_-\vr{z\downarrow} = 0$, where
$\vr{z\uparrow}$ and $\vr{z\downarrow}$ are spin states with a definite projection
on the $z$-axis. One easily observes the similarity of the operators $\sigma_\pm$
with the creation and annihilation operators, precisely we can set:
\be
\vr{0} = \vr{z\uparrow}, \quad \vr{1} = \vr{z\downarrow}, \quad a = \sigma_+,\quad
a^\d = \sigma_-.
\en{EJ12}
Then, for instance, $\sigma_z = 1-2\hat{n}$ by comparison of their actions on the
basis states.

The above local construction, however, is invalid for the set of the spin operators
$\sigma_\alpha^{(j)}$, $j=1,\ldots,s$,  since the different spin operators commute,
whereas the different fermion creation and annihilation operators satisfy the
anti-commutation relations. The required correspondence  between the two sets of
operators must be   \textit{non-local}. To arrive at the required correspondence,
let us  recall Eqs. (\ref{EJ8}) and (\ref{EJ9}). Using that the occupation number
$n_j$ corresponds to spin up $n_j=0$ or spin down $n_j=1$ state, that the different
fermion operators $\hat{n}_j$ commute and that $1-2n_j = (-1)^{n_j}$ we can define
the set of fermion operators for the set of the spin variables as follows
\be
a_j = \left(\prod\limits_{l=1}^{j-1}\sigma_z^{(l)}\right) \sigma_+^{(j)},\quad
a^\d_j = \left(\prod\limits_{l=1}^{j-1}\sigma_z^{(l)}\right) \sigma_-^{(j)}.
\en{EJ13}
This definition is consistent with the definition of the $\sigma^{(j)}_z$ that we
have started with, indeed, from Eq. (\ref{EJ13}) we derive
\be
\sigma^{(j)}_z  = \sigma^{(j)}_+\sigma^{(j)}_- -\sigma^{(j)}_-\sigma^{(j)}_+=
a^\d_j a_j - a_j a^\d_j = 1-2\hat{n}_j.
\en{EJ14}
Thus, Eq. (\ref{EJ13}) can be inverted to obtain $\sigma^{(j)}_\pm$:
\be
\sigma^{(j)}_+ = \left[\prod\limits_{l=1}^{j-1}(1-2\hat{n}_l)\right]a_j,\quad
\sigma^{(j)}_- = \left[\prod\limits_{l=1}^{j-1}(1-2\hat{n}_l)\right]a^\d_j.
\en{EJ15}
The other spin operators are easily derived using Eqs. (\ref{EJ14}) and
(\ref{EJ15}). We have
\be
\sigma^{(j)}_x
=\left[\prod\limits_{l=1}^{j-1}(1-2\hat{n}_l)\right](a_j+a^\d_j),\quad
\sigma^{(j)}_y =i\left[\prod\limits_{l=1}^{j-1}(1-2\hat{n}_l)\right](a^\d_j-a_j).
\en{EJ16}

In the models of interacting planar spin systems, one usually encounters the terms
like the product of the neighboring spin operators, for instance,
$\sigma^{(j)}_x\sigma^{(j+1)}_x$ or $\sigma^{(j)}_y\sigma^{(j+1)}_y$. These
products simplify to the following  expressions \textit{quadratic} in  the fermion
operators
\be
\sigma^{(j)}_x\sigma^{(j+1)}_x =(a^\d_{j+1}+a_{j+1})(a_j - a^\d_j),\quad
\sigma^{(j)}_x\sigma^{(j+1)}_y =(a^\d_{j+1}-a_{j+1})(a^\d_j + a_j).
\en{EJ17}
Similar relations are valid for the cross-products of the spin components in the
$xy$ plane:
\be
\sigma^{(j)}_x\sigma^{(j+1)}_y = i(a^\d_j - a_j)(a^\d_{j+1}-a_{j+1}),\quad
\sigma^{(j)}_x\sigma^{(j+1)}_y = i(a^\d_{j+1}+a_{j+1})(a^\d_j + a_j).
\en{EJ18}
Therefore, any planar  spin model with the next-neighbor interaction between the
spins placed in the orthogonal magnetic field, i.e. describing the interacting spin
components in the plane perpendicular to the magnetic field, can be mapped by the
Jordan-Wigner transformation to a system of fermions described by a quadratic
Hamiltonian, which then can be diagonalized by a canonical transformation of
section \ref{secII1}. For example, consider the $XY$ spin chain with the
Hamiltonian
\be
H_{XY} =  \sum_{j}\alpha_j\sigma^{(j)}_z +
\sum_{j}\beta_j\sigma^{(j)}_x\sigma^{(j+1)}_x +
\sum_j\gamma_j\sigma^{(j)}_y\sigma^{(j+1)}_y,
\en{EJ19}
where $\alpha$ is proportional to the external magnetic field along the $z$-axis
and the coefficients $\beta$ and $\gamma$ describe the spin interaction along the
respective axes. In the fermion variables the spin Hamiltonian becomes
\[
H_{XY} =\frac12\sum_{j}\left( \alpha_j (a^\d_j a_j - a_j a^\d_j) +
\frac{\beta_j+i\gamma_j}{2}a^\d_{j+1}a_j +\frac{\beta_j-i\gamma_j}{2}a^\d_j
a_{j+1}\right)
\]
\be
+  \frac12\sum_j \frac{\beta_j+i\gamma_j}{2}a_{j+1}a_j
-\frac12\sum_j\frac{\beta_j-i\gamma_j}{2} a^\d_{j+1}a^\d_j.
\en{EJ20}


\begin{thebibliography}{99}


\bibitem{BgBook} N.N. Bogoliubov, N.N. Bogoliubov Jr.,
\textit{Introduction to Quantum Statistical Mechanics} (World Scientific,
Singapore, 1982).

\bibitem{PO} O. Penrose and L. Onsager, Phys. Rev. \textbf{104,} 576 (1956).

\bibitem{LRO} C. N. Yang, Rev. Mod. Phys. \textbf{34,} 694 (1962).

\bibitem{BgMeth} N. N. Bogoliubov, Izv. Akad. Nauk SSSR \textbf{11,} 77 (1947);
J. Phys.  USSR \textbf{11,} 23 (1947).

\bibitem{BgZub} N. N. Bogoliubov and D. N. Zubarev, Soviet Phys. [JETP] \textbf{1,}
83 (1955).

\bibitem{PSBook} L. P. Pitaevskii and S. Stringari, \textit{Bose-Einstein Condensates in
Gases} (Cambridge University Press, Cambridge, England, 2003).



\bibitem{BgAppr} E. H. Lieb, R. Seiringer and J. Yngvason, Phys. Rev. Lett.
\textbf{94,} 080401 (2005); A. S\"ut\H{o}, Phys. Rev. Lett. \textbf{94,} 080402
(2005).


\bibitem{CD-BEC} Y. Castin and R. Dum, Phys. Rev. A \textbf{57,} 3008 (1998).

\bibitem{NPBook} Ph. Nozi\`eres and D. Pines, \textit{The Theory  of Quantum
Liquids} (Addison-Wesley, New York, 1990).

\end{thebibliography}
\end{document}